\documentclass[journal]{IEEEtran}
\usepackage{amsthm}
\usepackage{placeins}
\usepackage{dsfont}
\usepackage{amsfonts}
\usepackage{amssymb}
\usepackage{mathrsfs}
\usepackage{mathtools}
\usepackage[english]{babel}
\usepackage{float}
\usepackage[pdftex]{graphicx}
\usepackage{epstopdf}
\usepackage{amsmath}
\usepackage{color}
\usepackage{multirow}
\usepackage[switch]{lineno}
\usepackage[named]{algo}
\usepackage{algpseudocode}
\usepackage{algorithm}
\usepackage{algorithmicx}
\usepackage{bm}
\usepackage{balance}
\usepackage{stfloats}
\usepackage{lipsum}
\usepackage[inline]{enumitem}
\usepackage{microtype}
\usepackage{cuted}
\usepackage[caption=false,font=footnotesize]{subfig}
\usepackage{cite}
\usepackage{hyperref}
\usepackage{scalerel}
\usepackage{tikz}
\usetikzlibrary{decorations.pathreplacing}
\usetikzlibrary{arrows}
\usepackage{siunitx}
\DeclareMathAlphabet\mathbfcal{OMS}{cmsy}{b}{n}
\newtheorem{theo}{Theorem}
\newtheorem{lem}{Lemma}

\newtheorem{Col}{Corollary}
\theoremstyle{definition}
\newtheorem{rmk}{Remark}
\newtheorem{Def}{Definition}

\newcommand{\EX}{{\mathbb{E}}}           
\newcommand{\ZEROV}{{\mathbf{0}}}        
\newcommand{\IDM}{{\mathbf{I}}}          
\newcommand{\DIAG}{{\mathrm{diag}}}      
\newcommand{\TRAC}{{\mathrm{tr}}}        
\newcommand{\VE}{{\mathrm{vec}}}         
\newcommand{\MAT}{{\mathrm{mat}}}         
\newcommand{\RK}{{\mathrm{rank}}}        
\newcommand{\SGN}{{\mathrm{sgn}}}        
\newcommand{\asin}{{\mathrm{arcsine}}}        
\newcommand{\sine}{{\mathrm{sine}}}        
\newcommand{\J}{j}

\newcommand{\AM}{\mathbf{A}}
\newcommand{\CM}{\mathbf{C}}
\newcommand{\EM}{\mathbf{E}}
\newcommand{\FM}{\mathbf{F}}

\newcommand{\GM}{\mathbf{G}}
\newcommand{\JM}{\mathbf{J}}
\newcommand{\KM}{\mathbf{K}}
\newcommand{\LM}{\mathbf{L}}
\newcommand{\MM}{\mathbf{M}}
\newcommand{\QM}{\mathbf{Q}}
\newcommand{\RM}{\mathbf{R}}
\newcommand{\TM}{\mathbf{T}}
\newcommand{\UM}{\mathbf{U}}
\newcommand{\VM}{\mathbf{V}}
\newcommand{\WM}{\mathbf{W}}
\newcommand{\XM}{\mathbf{X}}
\newcommand{\YM}{\mathbf{Y}}

\newcommand{\AV}{\mathbf{a}}
\newcommand{\BV}{\mathbf{b}}
\newcommand{\CV}{\mathbf{c}}
\newcommand{\DV}{\mathbf{d}}
\newcommand{\EV}{\mathbf{e}}
\newcommand{\GV}{\mathbf{g}}
\newcommand{\HV}{\mathbf{h}}
\newcommand{\NV}{\mathbf{n}}
\newcommand{\PV}{\mathbf{p}}

\newcommand{\RV}{\mathbf{r}}
\newcommand{\SV}{\mathbf{s}}

\newcommand{\UV}{\mathbf{u}}
\newcommand{\VV}{\mathbf{v}}

\newcommand{\XV}{\mathbf{x}}
\newcommand{\YV}{\mathbf{y}}
\newcommand{\ZV}{\mathbf{z}}
\newcommand{\PHI}{\boldsymbol{\phi}}
\newcommand{\VAREPSILON}{\boldsymbol{\varepsilon}}
\newcommand{\VARRHO}{\boldsymbol{\varrho}}
\newcommand{\IMATH}{\boldsymbol{\imath}}

\newcommand{\THETA}{\boldsymbol{\theta}}
\newcommand{\GAMMA}{\boldsymbol{\gamma}}
\newcommand{\ALPHA}{\boldsymbol{\alpha}}
\newcommand{\BETA}{\boldsymbol{\beta}}
\newcommand{\PHIM}{\mathbf{\Phi}}
\newcommand{\SIGMAM}{\mathbf{\Sigma}}
\newcommand{\PSIM}{\mathbf{\Psi}}
\newcommand{\GAMMAM}{\mathbf{\Gamma}}
\newcommand{\UPSILONM}{\mathbf{\Upsilon}}
\newcommand{\DELTAM}{\mathbf{\Delta}}

\newcommand{\OMEGAM}{\mathbf{\Omega}}
\newcommand{\DIGAMMAM}{\mathbf{\digamma}}
\newcommand*{\Scale}[2][4]{\scalebox{#1}{$#2$}}

\makeatletter
\IEEEtriggercmd{\reset@font\normalfont\fontsize{7pt}{7.8pt}\selectfont}
\makeatother
\IEEEtriggeratref{1}
\IEEEpubid{\begin{minipage}[t]{\textwidth}\ \\[10pt]
\centering\scriptsize{\textcopyright~2021 IEEE. Personal use of this material is permitted. Permission from IEEE must be obtained for all other uses, in any current or future media, including reprinting/republishing this material for advertising or promotional purposes, creating new collective works, for resale or redistribution to servers or lists, or reuse of any copyrighted component of this work in other works.}
\end{minipage}}
\begin{document}

\title{On the Performance of One-Bit DoA Estimation via Sparse Linear Arrays}

\author{\IEEEauthorblockN{Saeid~Sedighi,~\IEEEmembership{Student Member,~IEEE,} M.~R.~Bhavani~Shankar,~\IEEEmembership{Senior Member,~IEEE,} Mojtaba Soltanalian,~\IEEEmembership{Senior Member,~IEEE,} and Bj\"{o}rn Ottersten,~\IEEEmembership{Fellow,~IEEE}\\}
\thanks{
This work was supported in part by the Luxembourg National Research Fund (FNR) under the ACCORDION
(ref: $11228830$) and SPRINGER (ref: 12734677) projects, the European Research Council (ERC) Grant AGNOSTIC (ID: 742648), and U.S. National Science Foundation (NSF) Grants $1704401$, $1809225$, and an Illinois Discovery Partners Institute  (DPI) Seed Award.

S. Sedighi, M. R. B. Shankar and B. Ottersten are with the Interdisciplinary Centre for Security, Reliability and Trust (SnT), University of Luxembourg, Luxembourg City L-1855, Luxembourg (e-mails: saeid.sedighi@uni.lu; bhavani.shankar@uni.lu; bjorn.ottersten@uni.lu). M. Soltanalian is with the Department of Electrical and Computer Engineering, University of Illinois at Chicago, Chicago, IL 60607 USA (e-mail: msol@uic.edu).

This paper has supplementary downloadable material available at http://ieeexplore.ieee.org, provided by the author. 
The material includes Appendix K. This material is $183$ KB in size.
}}
\maketitle
\begin{abstract}
Direction of Arrival (DoA) estimation using Sparse Linear Arrays (SLAs) has recently gained considerable attention in array processing thanks to their capability to provide enhanced degrees of freedom in resolving uncorrelated source signals. Additionally, deployment of one-bit Analog-to-Digital Converters (ADCs) has emerged as an important topic in array processing, as it offers both a low-cost and a low-complexity implementation. In this paper, we study the problem of DoA estimation from one-bit measurements received by an SLA. Specifically, we first investigate the identifiability conditions for the DoA estimation problem from one-bit SLA data and establish an equivalency with the case when DoAs are estimated from infinite-bit unquantized measurements. Towards determining the performance limits of DoA estimation from one-bit quantized data, we derive a pessimistic approximation of the corresponding Cram\'{e}r-Rao Bound (CRB). This pessimistic CRB is then used as a benchmark for assessing the performance of one-bit DoA estimators. We also propose a new algorithm for estimating DoAs from one-bit quantized data. We investigate the analytical performance of the proposed method through deriving a closed-form expression for the covariance matrix of the asymptotic distribution of the DoA estimation errors and show that it outperforms the existing algorithms in the literature. Numerical simulations are provided to validate the analytical derivations and corroborate the resulting performance improvement.
\end{abstract}


\IEEEpeerreviewmaketitle

\section{Introduction}
The problem of Direction of Arrival (DoA) estimation is of central
importance in the field of array processing with many applications
in radar, sonar, and wireless communications \cite{van2002optimum,Haykin1992,Bjorn1996}. Estimating DoAs using Uniform Linear Arrays (ULAs) is well-investigated in the literature; a number of algorithms such as the Maximum Likelihood (ML) estimator, MUSIC, ESPRIT and subspace
fitting were presented and their performance thoroughly analyzed \cite{paulraj1993,Li1993,stoica1989,Stoica19901,Viberg1991,Jansson1999}. However, it is widely
known that ULAs are not capable of identifying more sources than the number of physical elements in the array \cite{Haykin1992,Stoica19901}.

To transcend this limitation, exploitation of Sparse Linear Arrays (SLAs) with particular geometries, such as Minimum Redundancy Arrays (MRAs) \cite{Moffet1968}, co-prime
arrays \cite{Vaidyanathan2011} and nested arrays \cite{Pal2010} has been proposed. These architectures can dramatically boost the degrees of freedom of the array
for uncorrelated source signals such that a significantly larger number of
sources than the number of physical elements in the array can be identified. In addition, the enhanced degrees of freedom provided by these SLAs can improve the resolution performance appreciably compared to ULAs \cite{Pal2010}.
These features have spurred further research on DoA estimation using SLAs in recent years.
A detailed study on DoA estimation via SLAs through an analysis of the Cram\'{e}r-Rao Bound (CRB) was conducted in \cite{Liu2017}.
Further, a number of approaches to estimating DoAs from SLA measurements were proposed in the literature. In general, the proposed approaches can be classified under two main groups: \begin{enumerate*} \item Sparsity-Based Methods (SBMs); \item Augmented Covariance-Based Methods (ACBMs)\end{enumerate*}. SBMs estimate DoAs by imposing sparsity constraints on source profiles and exploiting the compressive sensing recovery techniques \cite{Zhang2013,Shen2016,Pal2015,Pal2012con,Chi2010,Tan2014,Yang2014}.
However, in ACBMs, DoAs are estimated by applying conventional subspace methods such as MUSIC and ESPRIT on an Augmented Sample Covariance Matrix (ASCM) developed from the original sample covariance matrix by exploiting the difference co-array structure~\cite{Pal2010,Wang2017,Sedighi2018SAM}. In addition, the authors of this paper recently proposed a Weighted Least Squares (WLS) estimator capable of asymptotically achieving the corresponding CRB for DoA estimation from SLA data \cite{SedighiTSP2019,Sedighi2018Asilomar}.

The aforementioned techniques for DoA estimation from SLA data rest on the assumption that the analog array measurements are digitally represented by a significantly large number of bits per sample such that the resulting quantization errors can be disregarded. However, the production costs and energy consumption of Analog-to-Digital Converters (ADCs) escalate dramatically as the number of quantization bits and sampling rate increase \cite{Walden1999}. In consequence, deployment of high-resolution ADCs in many modern applications, e.g. cognitive radio \cite{Sun2013}, cognitive radars \cite{Lunden2015}, automotive radars \cite{Hasch2012}, radio astronomy\cite{burke2019introduction} and massive multiple-input multiple-output (MIMO) systems\cite{Lu2014}, is not economically viable owing to their very high bandwidth.
In order to reduce energy consumption and production cost in such applications, researchers and system designers have recently proposed using low-resolution ADCs. As an extreme case of low-resolution ADCs,
one-bit ADCs, which convert an analog signal into digital data using a single bit per sample, has received significant attention in the literature. One-bit ADCs offer an extremely high sampling rate at a low cost and very low energy consumption \cite{Walden1999}. Additionally, they enjoy the benefits of relatively easy implementation due to their simple architecture \cite{pelgrom2013analog}. In the past few years, numerous studies were conducted to investigate the impact of using one-bit sampling
on various applications such as massive MIMO systems \cite{Gokceoglu2017,Saxena2017,Rao2019,Pirzadeh2020,Wan2020},
dictionary learning \cite{zayyani2015dictionary},
radar \cite{zhao2018deceptive,ameri2019,Zahabi2020,Xi2020,sedighi2020localization}, and array processing \cite{BarShalom2002,Stein2016}.
\subsection{Relevant Works}
The problem of DoA estimation from one-bit quantized data has been studied in the literature presuming both the deterministic signal model \cite{stoica1989} and the stochastic signal model \cite{Stoica19901}. The studies in  \cite{Huang2020,Yoffe2019,Stockle2015,Huang2018,Meng2018} presuppose the deterministic signal model. The authors in \cite{Huang2020} developed an algorithm for reconstruction of the unquantized array measurements from one-bit samples followed by MUSIC to determine DOAs. The ML estimation was deployed in \cite{Yoffe2019} for finding DoAs from one-bit data. In \cite{Meng2018}, the authors utilized a sparse Bayesian learning algorithm to solve the DoA estimation problem from one-bit samples. Two sparsity-based approaches were also proposed in \cite{Stockle2015,Huang2018}. Further, DoA estimation from one-bit data assuming the stochastic signal model has been discussed in \cite{BarShalom2002,Stein2016,Chen2018,Huang2019}.
In the special case of a two-sensor array, the exact CRB expression for the DoA estimation problem from one-bit quantized data was derived in \cite{BarShalom2002}. Moreover, an approach for estimating DoAs from one-bit ULA samples was proposed in \cite{BarShalom2002} which is based on reconstruction of the covariance matrix of unquantized data using the arcsine law \cite{VanVleck1966}.
In contrast to the approach employed in \cite{Liuonebit} which relies on the covaraince matrix reconstruction of unquantized data, the DoA estimation was performed in \cite{Huang2019} by directly applying MUSIC on the sample covariance matrix of one-bit ULA data. The numerical simulations demonstrated that the approach proposed in \cite{Huang2019} performs similar to the algorithm proposed in \cite{BarShalom2002} in the low Signal-to-Noise Ratio (SNR) regime.
An upper bound on the CRB of estimating a single source DoA from one-bit ULA measurements was derived in \cite{Stein2016}.

The aforementioned research works considered using ULAs for one-bit DoA estimation. Exploitation of SLAs for one-bit DoA estimation has been studied in \cite{Liuonebit,Ramamohan2010,Cheng2020,Zhou2020}. The authors in \cite{Liuonebit} deployed the arcsine law \cite{VanVleck1966} to reconstruct the ASCM from one-bit SLA data. Then, they applied MUSIC on the reconstructed ASCM to estimate DoAs.
It was shown in \cite{Liuonebit} that the performance degradation due to one-bit quantization can,
to some extent, be compensated using SLAs. An array interpolation-based algorithm was employed in \cite{Zhou2020} to estimate DoAs from one-bit data received by co-prime arrays. Cross-dipoles sparse arrays were deployed in \cite{Cheng2020} to develop a method for one-bit DoA estimation which is robust against polarization states. In \cite{Ramamohan2010}, the authors proposed an approach to jointly estimate DoAs and array calibration errors from one-bit data.

Nonetheless, the analytical performance of DoA estimation from one-bit SLA measurements has not yet been studied in the literature and performance analysis in the literature has been limited to simulations studies. Therefore, fundamental performance limitations of DoA estimation form one-bit SLA measurements have not well understood.

\subsection{Our Contributions}
It is of great importance to analytically investigate the performance of DoA estimation from one-bit SLA measurements.
Such a performance analysis not only provides us with valuable insights into the performance of DoA estimation from one-bit SLA data but also enables us to compare its performance with that of DoA estimation using infinite-bit (unquantized) SLA data. Hence, as one of the contributions of this paper, we conduct a rigorous study on the performance of estimating source DoAs from one-bit SLA samples. Furthermore, we propose a new algorithm for estimating source DoAs from one-bit SLA measurements and analyze its asymptotic performance. Specifically, the contributions of this paper are described as follows:
\begin{itemize}
\item {\bf Identifiability Analysis:} We study the identifiability conditions for the DoA estimation from one-bit SLA data. We first show that the identifiability condition for estimating DoAs from one-bit SLA data is equivalent to the case when DoAs are estimated from infinite-bit (unquantized) SLA data. Then, we determine a sufficient condition for global identifiablity of DoAs from one-bit data based on the relationship between the number of source and array elements.

\item {\bf CRB Derivation and Analysis:} We derive a pessimistic approximation of the CRB of DoA estimation using one-bit data received by an SLA. This pessimistic CRB approximation provides a benchmark for the performance of DoA estimation algorithms from one-bit data. Additionally, it helps us to spell out the condition under which the Fisher Information Matrix (FIM) of one-bit data is invertible, and thus, the CRB is a valid bound for one-bit DoA estimators. Further, we derive the performance limits of one-bit DoA estimation using SLAs at different conditions.

\item {\bf Novel One-bit DoA Estimator:} We propose a new MUSIC-based algorithm for estimating DoAs from one-bit SLA measurements. In this regard, we first construct an enhanced estimate of the normalized covariance matrix of infinite-bit (unquantized) data by exploiting the structure of the normalized covariance matrix efficiently. Then, we apply MUSIC to an augmented version of the enhanced normalized covariance matrix estimate to determine the DoAs.

\item {\bf Performance Analysis of the Proposed Estimator:} We derive a closed-form
expression for the second-order statistics of the asymptotic distribution (for the large number of snapshots) of the proposed algorithm. Our asymptotic performance analysis shows that the proposed estimator outperform its counterparts in the literature and that its performance is very close to the proposed pessimistic approximation of the CRB. Moreover, the asymptotic performance analysis of the proposed DoA estimator enables us to provide valuable insights on its performance. For examples, we observe that the Mean Square Error (MSE) depends on both the physical array geometry and the co-array geometry. In addition, we observe that the MSE does not drop to zero even if the SNR approaches infinity.

\item  {\bf Wider Applicability of the derived performance Analysis:} We provide a closed-form expression for the large sample performance of the one-bit DoA estimator in \cite{Liuonebit} as a byproduct of the performance analysis of our proposed DoA estimator.
\end{itemize}

\emph{Organization}: Section \ref{sec:model} describes the system model.
In Section \ref{sec:iden}, the identifiability condition for DoA estimation problem from one-bit quantized data is discussed. Section \ref{sec:crb} presents the pessimistic approximation of the CRB and related discussions.  In Section \ref{sec:est}, the proposed algorithm for DoA estimation from one-bit measurements is given and its performance is analyzed.
The simulation results and related discussions are included in Section \ref{sec:simulations}. Finally, Section \ref{sec:conclusion} concludes the paper.

\emph{Notation}: Vectors and matrices are referred to by lower- and upper-case bold-face, respectively. The superscripts $*$, $T$, $H$ denote the conjugate, transpose and Hermitian
(conjugate transpose) operations, respectively.
$[\AM]_{i,j}$ and $[\AV]_i$ indicate the $(i,j)^{\rm th}$ and $i^{\rm th}$ entry of $\AM$ and $\AV$, respectively.
$\|\AV\|_2$ stands for the $\ell_2$-norm of $\AV$. $|\mathds{A}|$ represents the cardinality of the set $\mathds{A}$. $|a|$, $\lceil a \rceil$ and $\lfloor a \rfloor$ represent the absolute value of, the least integer greater than or equal to and greatest integer less than or equal to the scalar $a$, respectively. $\DIAG(\AV)$ and $\DIAG(\AM)$ are
diagonal matrices whose diagonal entries are equal to the elements of $\AV$ and to the diagonal elements of $\AM$, respectively. The $M \times M$ identity matrix is denoted by $\mathbf{I}_M$. $\SGN(x)$ denotes the sign function with $\SGN(x) = 1$ for $x \geq 0$ and $\SGN(x) = - 1$ otherwise.
The real and image part
of $a$ are denoted by $\Re\{a\}$ and $\Im\{a\}$, respectively.
$\EX\{.\}$ stands for the statistical expectation.
$\otimes$ and $\odot$ represent Kronecker and Khatri-Rao products, respectively. $\TRAC(\AM)$, $\det(\AM)$ and $\RK(\AM)$
denote the trace, determinant and rank, respectively.
$
\VE\left(\AM\right)=
\begin{bmatrix}
\AV_1^T & \AV_2^T & \cdots & \AV_n^T \\
\end{bmatrix}^T
$
represents the vectorization operation and $\MAT_{m,n}(.)$ is its inverse operation.
$\mathbf{A}^{\dagger}$ and $\Pi^{\bot}_{\mathbf{A}}$ indicate the pseudoinverse and the projection matrix onto the null space of the full column rank matrix $\mathbf{A}^H$, respectively. ${\cal CN}(\AV, \AM)$ denote the circular complex Gaussian distribution with mean $\AV$ and covariance matrix $\AM$.

\section{System Model}\label{sec:model}
\begin{figure}[!t]
\centering
\begin{tikzpicture}[scale=0.4]
\draw[semithick] (-10,0) -- (10,0);
\draw[semithick] (-10,2) -- (10,2);
\foreach \x in {-9,-8,-7,-6,-5,-4,-3,-2,0,-1,1,2,3,4,5,6,7,8,9}
    \draw (\x cm, 5pt) -- (\x cm, -5pt) node[anchor=north] {\tiny $\x$};
\fill [radius=5pt,color=black] (-9,0) circle[] (-7,0) circle[] (-6,0) circle[] (-5,0) circle[] (-4,0) circle[] (-3,0) circle[] (-2,0) circle[] (-1,0) circle[] (0,0) circle[] (1,0) circle[] (2,0) circle[] (3,0) circle[] (4,0) circle[] (5,0) circle[] (6,0) circle[] (7,0) circle[] (9,0) circle;
\foreach \x in {-9,-8,-7,-6,-5,-4,-3,-2,0,-1,1,2,3,4,5,6,7,8,9}
    \draw (\x cm, 61.6929pt) -- (\x cm, 51.6929pt);
    \fill [radius=5pt,color=black] (0,2) circle[] (2,2) circle[] (3,2) circle[] (4,2) circle[] (6,2) circle[] (9,2) circle;
\node at (-11,2) {(a)};
\node at (-11,0) {(b)};
\draw [decorate,decoration={brace,amplitude=5pt,mirror,raise=3ex}]
  (-7,0.0) -- (7,0.0) node[midway,yshift=-2.5em]{\footnotesize The contiguous ULA segment};
\draw [decorate,decoration={brace,amplitude=2pt,raise=1ex}]
  (0,2) -- (1,2) node[midway,yshift=1.35em]{\footnotesize $\frac{\lambda}{2}$};
\end{tikzpicture}
\vspace*{-0.12in}
\DeclareGraphicsExtensions.
\caption{Array geometry of a co-prime array with $M=6$ elements: (a) physical array with $\mathds{M}=\{0,2,3,4,6,9\}$; (b) difference co-array with $\mathds{D}=\{0,1,2,3,4,5,6,7,9\}$ and $v=8$.}
\label{fig-1}
\vspace{-5 mm}
\end{figure}
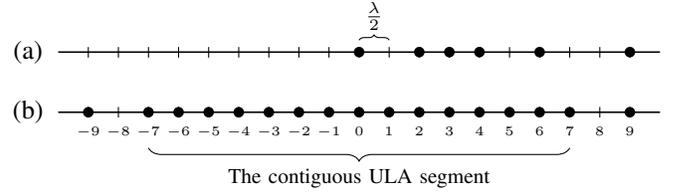

We consider an SLA with $M$ elements located at positions $m_1\frac{\lambda}{2}, m_2\frac{\lambda}{2}, \cdots, m_M\frac{\lambda}{2}$ with $m_i \in \mathds{M}$. Here $\mathds{M}$ is a set of integers with cardinality $|\mathds{M}|\!=\!M$, and $\lambda$ denotes the wavelength of the incoming signals. It is assumed that $K$ narrowband signals with distinct DoAs $\THETA\!=\![ \theta_1, \theta_2, \cdots, \theta_K ]^T \in [-\pi/2, \pi/2]^{K \times 1}$ impinge on the SLA from far field. The signal received at the array at time instance $t$ can be modeled as
\small
\begin{align}\label{model-eq-1}
\YV(t)=\AM(\THETA)\SV(t)+\NV(t) \in \mathds{C}^{M \times 1}, ~~~ t=0,\cdots, N-1,
\end{align}\normalsize
where $\SV(t) \in \mathds{C}^{K \times 1}$ denotes the vector of source signals, $\NV(t) \in \mathds{C}^{M \times 1}$ is additive noise, and $\AM(\THETA)=[\AV\left(\theta_1\right), \AV\left(\theta_2\right), \cdots, \AV\left(\theta_K\right)] \in \mathds{C}^{M \times K}$ represents the SLA steering matrix with
\small
\begin{align}\label{model-eq-2}
\AV(\theta_k)\!=\![e^{\J\pi \sin \theta_k m_1},~ e^{\J\pi \sin \theta_k m_2},~ \cdots,~ e^{\J\pi \sin \theta_k m_M}]^T,
\end{align}\normalsize
being the SLA manifold vector for the $i^{\rm th}$ signal.
Further, the following assumptions are made on source signals and noise:
\begin{itemize}
  \item[{\bf A1}] $\NV(t)$ follows a zero-mean circular complex Gaussian distribution with the covariance matrix $\EX\{\NV(t)\NV^H(t)\}\!=\!\sigma^2\IDM_M$.
  \item[{\bf A2}] The source signals are modeled as zero-mean \emph{uncorrelated} circular complex Gaussian random variables with covariance matrix $\EX\{\SV(t)\SV^H(t)\}=\DIAG(\PV)$ where $\PV=[ p_1, p_2, \cdots, p_K ]^T \in \mathds{R}_{>0}^{{K \times 1}}$ (i.e., $p_k > 0,~\forall k$).
  \item[{\bf A3}] Source and noise vectors are  mutually independent.
  \item[{\bf A4}] There is no temporal correlation between the snapshots, i.e., $\EX\{\NV(t_1)\NV^H(t_2)\}\!=\!\EX\{\SV(t_1)\SV^H(t_2)\}\!=\!\mathbf{0}$ if $t_1\neq t_2$. 
 \item[{\bf A5}] An exact knowledge of the number of sources is available.
\end{itemize}
Given {\bf A1} - {\bf A4}, the covariance matrix of $\YV(t)$ is expressed as
\small
\begin{align}\label{model-eq-3}
\RM &= \AM(\THETA)\DIAG(\PV)
\AM^H(\THETA)+\sigma^2\IDM_M  \in  \mathds{C}^{M \times M}.
\end{align}\normalsize
Vectorizing $\overline{\RM}$ leads to \cite{Liu2017,Wang2017,SedighiTSP2019}
\small
\begin{align}\label{model-eq-4}
\RV& \doteq \VE (\RM) = \left(\AM^*(\THETA) \odot \AM(\THETA)\right)\PV+\sigma^2\VE(\IDM_M),\nonumber\\
&=\JM\AM_d(\THETA)\PV+\sigma^2\JM\EV \in \mathds{C}^{M^2 \times 1},
\end{align}\normalsize
where $\AM_d(\THETA) \in \mathds{C}^{(2D-1) \times K}$ corresponds to the steering matrix of the difference co-array of the SLA whose elements are located at $(-\ell_{D-1} \frac{\lambda}{2}, \cdots, 0, \cdots, \ell_{D-1} \frac{\lambda}{2})$ with $\ell_i \in \mathds{D}=\{\tiny|m_p-m_q\tiny| : m_p, m_q \in \mathds{M}\}$
and $D=|\mathds{D}|$. Moreover, $\EV \in \{0,1\}^{(2D-1) \times 1}$ is a column vector with $[\EV]_i=\delta[i-D]$, and the selection matrix $\JM \in \{0,1\}^{M^2 \times (2D-1)}$ is represented as follows \cite{Liu2017}: 
\small
\begin{align}\label{model-eq-5}
\JM \!=\! \begin{bmatrix} \VE (\LM^T_{D-1}), \!&\! \cdots, \!&\! \VE (\LM_{0}), \!&\! \cdots, \!&\! \VE (\LM_{D-1}) \end{bmatrix},
\end{align}\normalsize
where
$
\Scale[0.9]{[\LM_n]_{p,q}=\left\{\begin{array}{cc}
           1, & \text{if}~~ m_p-m_q=\ell_n, \\
            0, & \text{otherwise},
          \end{array}
\right.}
$
with $1 \leq p, q \leq M$ and $0 \leq n \leq D-1$. The steering matrix of the difference co-array includes a contiguous ULA segment around the origin with the size of $2v-1$ where $v$ is the largest integer such that $\{0, 1, \cdots, v-1\} \subseteq \mathds{D}$. The size of the contiguous ULA segment of the
difference co-array plays a crucial role in the number of identifiable sources such that $K$ distinct sources are identifiable if $K \leq v-1$. Hence, in case the SLA is designed properly
such that $v > M$, we are able to identify more sources than
the number of physical elements in the SLA; exploiting the resulting structure of $\RM$ efficiently\cite{Vaidyanathan2011,Pal2010,Liu2017,SedighiTSP2019}.
An illustrative example of an SLA, the corresponding difference co-array, and its contiguous ULA segment is presented in Fig. \ref{fig-1}.

Here it is assumed that each array element is connected to a one-bit ADC which directly converts the received analog signal into binary data by comparing the real and imaginary parts of the received signal individually with zero. In such a case, the one-bit measurements at the $m^{\rm th}$ array element are given by
\small
\begin{align}
\label{model-eq-6}
    \hspace{-1mm}[\XV(t)]_m \!=\! \frac{1}{\sqrt{2}} \SGN\left(\Re\{[\YV(t)]_m\} \right)
    \!+\! \frac{\J}{\sqrt{2}}~ \SGN\left(\Im\{[\YV(t)]_m \}\right).
\end{align}\normalsize
The problem under consideration is the estimation of source DoAs, i.e., $\THETA$, from one-bit quantized measurements, i.e., $\XM = \begin{bmatrix}\XV(0), & \XV(1), & \cdots, & \XV(N-1)\end{bmatrix}$, collected by the SLA.
\section{Identifiability Conditions}
\label{sec:iden}
Note that there is a significant information loss expected when going from infinite-bit (unquantized) data, i.e., $\YM=[\YV(0), \YV(1), \cdots, \YV(M)]$, to one-bit data, i.e., $\XM$. This information loss may affect the attractive capability of SLAs to identify a larger number of uncorrelated sources than the number of array elements. To address this concern,  we will consider the identifiability conditions for DoA estimation from one-bit SLA measurements in this section. Before proceeding further, we first need to give a clear definition of identifiability for this problem.
\begin{Def}[Identifiability]
\label{definition-1}
Let $f(\XM \!\mid\! \THETA, \PV, \sigma^2)$ denote the Probability Density Function (PDF) of $\XM$ parameterized by $\THETA$, $\PV$ and $\sigma^2$. Then, the source DoAs are said to be identifiable from $\XM$ at point $\THETA_0 \in [-\pi/2, \pi/2]^{K \times 1}$ if there exist no $\breve{\THETA} \neq \THETA_0 \in [-\pi/2, \pi/2]^{K \times 1}$ such that $f(\XM \mid \THETA_0, \PV, \sigma^2) = f(\XM \mid \breve{\THETA}, \breve{\PV}, \breve{\sigma}^2)$ for any arbitrary values of $\PV \in \mathds{R}_{>0}^{{K \times 1}}$, $\breve{\PV} \in \mathds{R}_{>0}^{{K \times 1}}$, $\sigma^2$ and $\breve{\sigma}^2$ \cite[Ch. 1, Definition 5.2]{lehmann2006theory} \cite[pp. 62]{van2000asymptotic}.
\end{Def}
\begin{rmk}
\label{rmk-1}
The above definition can be used for identifiabilty of $\THETA_0$ from $\YM$ by replacing $f(\XM \!\mid\! \THETA, \PV, \sigma^2)$ with $f(\YM \!\mid\! \THETA, \PV, \sigma^2)$.
\end{rmk}
Based on the above definition, the necessary and sufficient condition for a particular DoA point to be identifiable from one-bit SLA data is given in the following Theorem.
\begin{theo}
\label{Theo-1}
The source DoAs are identifiable from $\XM$ at $\THETA_0 \in [-\pi/2, \pi/2]^{K \times 1}$ if and only if they are identifiable from $\YM$ at $\THETA_0$.
\end{theo}
 \begin{proof}
See Appendix \ref{app-A}.
\end{proof}

The above Theorem shows that the identifiability condition for the DoA estimation problem from one-bit SLA measurements is equivalent to that for the DoA estimation problem from infinite-bit (unquantized) SLA measurements. Hence, the information loss arises from one-bit quantization does not influence the number of identifiable sources. However, Theorem \ref{Theo-1} simply spells out the identifiability condition of a single DoA point. A sufficient condition for global identifiablity of source DoAs from one-bit data is given in the following theorem.
\begin{Def}[Global identifiability]
\label{definition-2}
The source DoAs are said to be globally identifiable from $\XM$ if there exists no distinct $\THETA \in [-\pi/2, \pi/2]^{K \times 1}$ and $\breve{\THETA} \in [-\pi/2, \pi/2]^{K \times 1}$ such that $f(\XM \mid \THETA,\PV,\sigma^2) = f(\XM \mid \breve{\THETA},\breve{\PV},\breve{\sigma}^2)$ for any arbitrary values of $\PV \in \mathds{R}_{>0}^{{K \times 1}}$, $\breve{\PV} \in \mathds{R}_{>0}^{{K \times 1}}$, $\sigma^2$ and $\breve{\sigma^2}$.
\end{Def}
\begin{theo}
\label{Theo-2}
The sufficient conditions for global indentifiability and global non-indentifiability of source DoAs from one-bit SLA data are given as follows:
\begin{enumerate}
\item[{\bf S1}] The source DoAs are globally identifiable (with probability one) from $\XM$ for any value of $\THETA \in [-\pi/2, \pi/2]^{K \times 1}$ if $K \leq v-1$.
\item[{\bf S2}] The source DoAs are globally unidentifiable from $\XM$ for any value of $\THETA \in [-\pi/2, \pi/2]^{K \times 1}$ if $K \geq D$.
\end{enumerate}
\end{theo}
 \begin{proof}
See Appendix \ref{app-B}.
\end{proof}
Having revealed that one-bit quantization does not affect the indentifiability conditions of source DoAs, we will investigate the performance of DoA estimation from one-bit SLA data through a CRB analysis in the next section.
\section{Cram\'{e}r-Rao Bound Analysis}
\label{sec:crb}
It is well-known that the CRB offers a lower bound on the covariance of any unbiased estimator \cite{kay1993fundamentals}. Hence, it is considered as a standard metric for evaluating the performance of estimators. In particular, the CRB can
provide valuable insights into the fundamental limits of estimation for specific problems as well as the dependence of the estimation performance on various system parameters. Deriving a closed-form expression for the CRB requires knowledge of the data distribution. However, the data distribution may not be known for some problems. In such cases, the Gaussian assumption is a natural choice which leads to the largest (most pessimistic) CRB in a general class of
data distributions \cite{Stoica2011Lec}.

In the problem of DoA estimation from one-bit SLA measurements, the true PDF of one-bit data is obtained from the orthant probabilities \cite{abrahamson1964orthant} of Gaussian distribution, for which a closed-form expression is not available in general. Motivated by this fact, in what follows, we derive a pessimistic closed-form approximation for the CRB of the DoA estimation problem from one-bit SLA data through considering a Gaussian distribution for $\XV(t)$. This pessimistic closed-form approximation is used for benchmarking the performance of one-bit DoA estimators as well as for investigating the performance limits of the DoA estimation problem from one-bit data. Making use of assumptions {\bf A1}-{\bf A4}, it is readily confirmed that
$
\EX\{\XV(t)\} = \ZEROV.
$
Further, the arcsine law \cite{VanVleck1966} establishes the following relationship between $\RM$ and $\RM_{\XV}$:
\small
\begin{align}
\label{Eq-arclaw}
\RM_{\XV} = \EX\{\XV(t)\XV^H(t)\} = \frac{2}{\pi} \asin( \overline{\RM} ),
\end{align}\normalsize
where $[\asin(\overline{\RM})]_{m,n} \!=\! \arcsin(\Re\{[\overline{\RM}]_{m,n}\})\!+\!\J \arcsin(\Im\{[\overline{\RM}]_{m,n}\})$ and
\small
\begin{align}
\label{Eq-CRB-1}
\overline{\RM} \!=\! \frac{\RM}{\sigma^2+\sum_{k=1}^K p_k} \!=\! \AM(
\THETA) \DIAG(\overline{\PV}) \AM^H(
\THETA) \!+\! (1\!-\!\sum_{k=1}^K \overline{p}_k) \IDM_M,
\end{align}\normalsize
is the normalized covariance matrix of $\YV(t)$ with $\overline{\PV} = [\overline{p}_1, \overline{p}_2, \cdots, \overline{p}_K]^T$ and $\overline{p}_k = \frac{p_k}{\sigma^2+\sum_{k=1}^K p_k}$.
It follows from \eqref{Eq-arclaw} and \eqref{Eq-CRB-1} that $\RM_{\XV}$ is a function of the parameters $\THETA$ and $\overline{\PV}$. Let $\VARRHO = [\THETA, \overline{\PV}]^T$ denote the vector of unknown parameters. Then, considering the Gaussian assumption, the worst-case Fisher Information Matrix (FIM) ${\cal I}_{w}(\VARRHO)$ is given by \cite{kay1993fundamentals}
\small
\begin{align}
\label{Eq-CRB-2}
[{\cal I}_w(\VARRHO)]_{m,n} &= N \TRAC(\RM_{\XV}^{-1} \frac{\partial \RM_{\XV}}{\partial [\VARRHO]_m} \RM_{\XV}^{-1} \frac{\partial \RM_{\XV}}{\partial [\VARRHO]_n} )    \nonumber\\
& = N  \frac{\partial \RV_{\XV}^H}{\partial [\VARRHO]_m} (\RM_{\XV}^{-T} \otimes \RM_{\XV}^{-1}) \frac{\partial \RV_{\XV}}{\partial [\VARRHO]_n},
\end{align}\normalsize
where $\RV_{\XV}=\VE(\RM_{\XV})$ and the last equality is obtained by using the relation $\TRAC(\CM_1 \CM_2 \CM_3 \CM_4) = \VE^H(\CM_2^H) (\CM_1^T \otimes \CM_3 ) \VE(\CM_4)$. From \eqref{model-eq-4}, \eqref{Eq-arclaw} and \eqref{Eq-CRB-2}, we obtain
\footnotesize
\begin{align}
\label{Eq-CRB-3}
\hspace{-1mm} \RV_{\XV} \!=\! \frac{2}{\pi} \asin( \VE(\overline{\RM}) )\!=\! \frac{2}{\pi} \JM \asin\left(\AM_d(\THETA) \overline{\PV}\!+\!(1\!-\!\sum_{k=1}^K \overline{p}_k) \EV\right).
\end{align}\normalsize
Computing the derivative of $ \RV_{\XV}$ with respect to $\theta_k$ and $\overline{p}_k$ yields
\small

\begin{align}
\label{Eq-CRB-4}
\frac{\partial \RV_{\XV}}{\partial \theta_k} =& \J \pi \cos(\theta_k) \overline{p}_k \JM \DIAG(\DV) \\
&\times\left[ \DIAG(\overline{\HV}) \Re\{\AV_d(\theta_k)\} + \J \DIAG(\HV) \Im\{\AV_d(\theta_k)\}\right], \nonumber\\
\label{Eq-CRB-5}
\frac{\partial \RV_{\XV}}{\partial \overline{p}_k} =& \JM \left[ \DIAG(\HV) \Re\{\AV_d(\theta_k)\} + \J \DIAG(\overline{\HV}) \Im\{\AV_d(\theta_k)\}\right],
\end{align}\normalsize
where $\HV$ and $\overline{\HV}$ are given in \eqref{Eq-CRB-6} and \eqref{Eq-CRB-7} at the top of the next page, $\AV_d(\theta_k)$ denotes the $k^{\rm th}$ column of $\AM_d(\THETA)$ and $\DV=[-\ell_{D-1}, \cdots, \ell_0, \cdots, \ell_{D-1} ]^T$.
\begin{figure*}[!t]
\small
\begin{align}
\label{Eq-CRB-6}
\HV &= \begin{bmatrix} \frac{1}{\sqrt{1-|\Re\{\sum_{k=1}^K \overline{p}_k e^{-\J \pi \sin \theta_k \ell_{D-1}}\}|^2}} & \cdots & 0 & \cdots & \frac{1}{\sqrt{1-|\Re\{\sum_{k=1}^K \overline{p}_k e^{\J \pi \sin \theta_k \ell_{D-1}}\}|^2}}\end{bmatrix}^T,\\
\label{Eq-CRB-7}
\overline{\HV} &= \begin{bmatrix} \frac{1}{\sqrt{1-|\Im\{\sum_{k=1}^K \overline{p}_k e^{-\J \pi \sin \theta_k \ell_{D-1}}\}|^2}} & \cdots & 0 & \cdots & \frac{1}{\sqrt{1-|\Im\{\sum_{k=1}^K \overline{p}_k e^{\J \pi \sin \theta_k \ell_{D-1}}\}|^2}}\end{bmatrix}^T,
\end{align}\normalsize
\vspace{-1mm}
\hrulefill
\vspace{-2mm}
\end{figure*}
It follows from \eqref{Eq-CRB-2}, \eqref{Eq-CRB-4} and \eqref{Eq-CRB-5} that
\small
\begin{align}
    \label{Eq-CRB-8}
 {\cal I}_w(\VARRHO) = N \begin{bmatrix}
 \GM^H \\ \VM^H
 \end{bmatrix}  \JM^H (\RM_{\XV}^{-T} \otimes \RM_{\XV}^{-1}) \JM \begin{bmatrix}
 \GM & \VM
 \end{bmatrix},
\end{align}\normalsize
where
\small
\begin{align}
    \label{Eq-CRB-9}
 \GM =& \J \pi  \DIAG(\DV)  \big[ \DIAG(\overline{\HV}) \Re\{\AM_d(\THETA)\} \\
 &+ \J \DIAG(\HV) \Im\{\AM_d(\THETA)\}\big] \boldsymbol{\Phi}(\THETA) \DIAG(\overline{\PV}), \nonumber\\
 \label{Eq-CRB-10}
 \VM =& \DIAG(\HV) \Re\{\AM_d(\THETA)\} + \J \DIAG(\overline{\HV}) \Im\{\AM_d(\THETA)\},
\end{align}\normalsize
with $\boldsymbol{\Phi}(\THETA)=\DIAG([ \cos \theta_1, \cos \theta_2, \cdots, \cos \theta_K ]^T)$. If ${\cal I}_w(\VARRHO)$ is non-singular, a pessimistic approximation for the CRB of estimating DoAs from one-bit SLA data can be obtained through inverting ${\cal I}_w(\VARRHO)$. Hence, we need to first establish the non-singularity of ${\cal I}_w(\VARRHO)$.
\begin{lem}
\label{Lem-1}
Define $\UPSILONM = \begin{bmatrix} \DELTAM & \DIGAMMAM  \end{bmatrix} \in \mathds{C}^{(2D-1) \times 2K}$, where
\small
\begin{align}
 \label{Eq-CRB-11}
\DELTAM &= \DIAG(\DV)  \big[ \DIAG(\overline{\HV}) \Re\{\AM_d(\THETA)\} + \J \DIAG(\HV) \Im\{\AM_d(\THETA)\}\big],\\
\DIGAMMAM &= \DIAG(\HV) \Re\{\AM_d(\THETA)\} + \J \DIAG(\overline{\HV}) \Im\{\AM_d(\THETA)\}.
\end{align}\normalsize
Then, ${\cal I}_w(\VARRHO)$ is non-singular if and only if $\UPSILONM$ is full-column rank.
\end{lem}
 \begin{proof}
See Appendix \ref{app-C}
\end{proof}
\begin{rmk}
\label{rmk-2}
Assuming ${\cal I}(\varrho)$ to be the true FIM, it follows from ${\cal I}(\varrho) \succeq {\cal I}_w(\VARRHO)$ that
$\UPSILONM$ being full-column rank
is also a sufficient condition for the non-sigularity of ${\cal I}(\varrho)$.
\end{rmk}
\begin{theo}
\label{Theo-3}
Let $CRB(\THETA)$ denote the CRB for source DoAs $\THETA$ from $\XM$. If ${\cal I}_w(\VARRHO)$ is non-singular, then a pessimistic approximation of $CRB(\THETA)$, denoted by $CRB_w(\THETA)$, is given by
\small
\begin{align}
\label{Eq-Pes-CRB}
CRB(\THETA) \preceq CRB_w(\THETA) = \frac{1}{4 N \pi^2} (\QM^H \Pi_{\MM^{\frac{1}{2}} \VM}^{\perp} \QM)^{-1},
\end{align}\normalsize
where $\OMEGAM = \frac{1}{\pi}\GM$,
\vspace{-1mm}
\small
\begin{align}
\MM & = \JM^H \left(\asin(\overline{\RM}^T) \otimes \asin(\overline{\RM}) \right)^{-1} \JM,\\
\QM & = \MM^{\frac{1}{2}} \DIAG(\DV) \OMEGAM \PHIM(\THETA) \DIAG(\overline{\PV}),
\end{align}\normalsize
with $\GM$ and $\VM$ being given in \eqref{Eq-CRB-9} and \eqref{Eq-CRB-10}, respectively.
\end{theo}
\vspace{-1mm}
 \begin{proof}
See Appendix \ref{app-D}
\end{proof}
\begin{rmk}
\label{rmk-new-new-2}
We note that $CRB_w(\THETA)$ bears a superficial resemblance to the CRB expression for DoA estimation from unquantized data, given by \cite[Theorem 2]{Liu2017}
\begin{align}
CRB_{I}(\THETA) = \frac{1}{4 N \pi^2} (\widetilde{\QM}^H \Pi_{\widetilde{\MM}^{\frac{1}{2}} \widetilde{\VM}}^{\perp} \widetilde{\QM})^{-1},
\end{align}
where
\small
\begin{align}
\widetilde{\MM} & = \JM^H \left(\overline{\RM}^T \otimes \overline{\RM} \right)^{-1} \JM,\\
\widetilde{\QM} & = \widetilde{\MM}^{\frac{1}{2}} \DIAG(\DV) \AM_d(\THETA) \PHIM(\THETA) \DIAG(\overline{\PV}),\\
\widetilde{\VM} &= \begin{bmatrix} \AM_d(\THETA) & \EV \end{bmatrix}.
\end{align}\normalsize
\end{rmk}
%
%
\begin{theo}
\label{Theo-5}
Assume all sources have equal power $p$ and $SNR=p/\sigma^2$. Then, we have
\small
\begin{align}
\lim_{SNR \to \infty} CRB_w(\THETA) \succ \ZEROV.
\end{align}
\end{theo}\normalsize
\begin{proof}
See Appendix \ref{app-E}
\end{proof}
\begin{rmk}
Theorem \ref{Theo-5} implies that the $CRB_w(\THETA)$ does not go to zero as the SNR increases. As a consequence, in the one-bit DoA estimation problem, we may not be able to render estimation errors arbitrarily small by increasing the SNR.
\end{rmk}
\section{Proposed One-Bit DoA Estimator}
\label{sec:est}
In this section, we first derive an enhanced estimate of the normalized covariance matrix of $\YV(t)$, i.e., $\overline{\RM}$, from one-bit SLA measurements through exploiting the structure of $\overline{\RM}$. Then, we obtain DoA estimates by applying Co-Array-Based MUSIC (CAB-MUSIC) \cite{Liu2015,Wang2017} to the enhanced estimate of $\overline{\RM}$. Further, we investigate the analytical performance of the proposed method for estimating DoAs from one-bit measurements.
\vspace{-3mm}
\subsection{Enhanced One-Bit Co-Array-Based MUSIC}\label{sec:OBCABM}
It is deduced from the strong law of large numbers \cite[ch. 8]{papoulis1991stochastic} that the sample covariance matrix of one-bit data provides a consistent estimate of $\RM_{\XV}$ with probability $1$, i.e.,
$
\label{Eq-Est-1}
{\rm Pr}\left(\lim_{N \to \infty} \widehat{\RM}_{\XV} = \RM_{\XV}\right) =1
$,
where $\widehat{\RM}_{\XV} = \frac{1}{N} \XM \XM^H$. In addition,
reformulating \eqref{Eq-arclaw} gives $\overline{\RM}$ based on the covariance matrix of one-bit data as follows:
\small
\begin{align}
\label{Eq-Est-2}
\overline{\RM} =
 \sine ( \frac{\pi}{2} \RM_{\XV}),
\end{align}\normalsize
where $[\sine ( \frac{\pi}{2} \RM_{\XV})]_{m,n} = \sin(\frac{\pi}{2}\Re\{[\overline{\RM}]_{m,n}\}) + \J \sin(\frac{\pi}{2}\Im\{[\overline{\RM}]_{m,n}\})$.
Accordingly, a consistent estimate of $\overline{\RM}$ is obtained as
\small
\begin{align}
\label{Eq-Est-3}
\widetilde{\overline{\RM}} =
 \sine ( \frac{\pi}{2} \widehat{\RM}_{\XV}).
\end{align}\normalsize
Most of the algorithms in the literature employ $\widetilde{\overline{\RM}}$ for estimating DoAs from one-bit measurements \cite{BarShalom2002,Liuonebit}. However, an enhanced estimate of $\overline{\RM}$
compared to $\widetilde{\overline{\RM}}$ can be found if the structure of $\overline{\RM}$ is taken into account. This enhanced estimate could in turn yield a better DoA estimation performance. In what follows, we introduce such an enhanced estimate of $\overline{\RM}$ by exploiting its structure. Then, we use this enhanced estimate to improve the DoA estimation performance from one-bit data.

It is readily known from \eqref{Eq-CRB-1} that $\overline{\RM}$ has the following structure
\small
\begin{align}
\label{Eq-Est-4}
 \overline{\RM} = \IDM_M + \sum_{n=1}^{D-1} u_n \LM_n + \sum_{n=1}^{D-1} u_n^* \LM_n^T   ,
\end{align}\normalsize
where $u_n = \sum_{k=1}^K \overline{p}_k e^{\J\pi \sin \theta_k \ell_n}$ and $\LM_n$ is given after eq. \eqref{model-eq-5} for $1 \leq n \leq D-1$. It can be observed from \eqref{Eq-Est-4} that the diagonal elements of $\overline{\RM}$ are all one while the off-diagonal elements are parameterized by the vector $\UV = [u_1, \cdots, u_{D-1}]^T \in \mathds{C}^{(D-1) \times 1}$. This means that there exist only $2D-2$ free real parameters in $\overline{\RM}$. Let $\ddot{\RV} \in \mathds{C}^{(M^2-M) \times 1}$ be the vector containing the off-diagonal elements of $\overline{\RM}$, obtained by removing the diagonal elements of $\overline{\RM}$ from $\VE(\overline{\RM})$. Evidently, $\ddot{\RV}$ is given by
\small
\begin{align}
\label{Eq-Est-5}
\ddot{\RV} &= \overline{\JM} \begin{bmatrix} \UV^* & \UV \end{bmatrix}^T = \overline{\JM} \PSIM \PHI,
\end{align}\normalsize
where $\PHI = [\Re\{\UV\}^T,  \Im\{\UV\}^T]^T \in \mathds{R}^{(2D-1 \times 1)}$,
\small
\begin{align}
\label{Eq-Est-PSIM}
 \PSIM = \begin{bmatrix}
 \IDM_{D-1} & -\J \IDM_{D-1} \\
 \IDM_{D-1} & \J \IDM_{D-1}
 \end{bmatrix} .
\end{align}\normalsize
and $\overline{\JM} \in \{0,1\}^{(M^2-M) \times (2D-2)}$
is obtained by removing the $D$-th column as well as the rows with indices $(i-1)M+1$ for all $1 \leq i \leq M$ from $\JM$.
%
%
It follows from \eqref{Eq-Est-5} that $\overline{\RM}$ is parameterized by the real-valued vector $\PHI$.
We wish to find $\PHI \in \mathds{E}_{\PHI} = \{ \PHI \mid \overline{\RM}(\PHI) \succeq \ZEROV\}$ from $\widehat{\RM}_{\XV}$. To this end,
let $\widehat{\ddot{\RV}}_{\XV} \in \mathds{R}^{(M^2-M) \times 1}$ denote the vector containing the off-diagonal elements of $\widehat{\RM}_{\XV}$, obtained by removing the diagonal entries of $\widehat{\RM}_{\XV}$ from $\VE(\widehat{\RM}_{\XV})$.
For large $N$, it follows from the Central Limit Theorem (CLT) \cite[ch. 8]{papoulis1991stochastic} that the distribution of $\widehat{\ddot{\RV}}_{\XV}$ asymptotically approaches a complex proper Gaussian distribution, i.e., $\widehat{\ddot{\RV}}_{\XV} \stackrel{D}{\rightarrow} {\cal CN}(\ddot{\RV}_{\XV}, \frac{4}{\pi^2 N}\SIGMAM)$, where $\ddot{\RV}_{\XV}$ is the vector obtained from stacking the off-diagonal elements of $\RM_{\XV}$ and $\SIGMAM = \frac{\pi^2 N}{4} \EX\{(\widehat{\ddot{\RV}}_{\XV}-\ddot{\RV}_{\XV}) (\widehat{\ddot{\RV}}_{\XV}-\ddot{\RV}_{\XV})^H\} \in \mathds{C}^{(M^2 -M) \times (M^2-M)}$. The closed-form expressions for the elements of $\SIGMAM$ are provided in Appendix K (kindly refer to the supplementary document). It is observed that the elements of $\SIGMAM$ are functions of $\ddot{\RV}$, thereby parameterized by $\PHI$ as well. Considering the transformation \eqref{Eq-Est-3}, the asymptotic distribution of the off-diagonal elements of $\widetilde{\overline{\RM}}$, denoted by $\widetilde{\ddot{\RV}} \in \mathds{C}^{(M^2-M) \times 1}$, is given by \eqref{Eq-Est-7} at top of this page.
\begin{figure*}[!t]
\small
\begin{align}
\label{Eq-Est-7}
&f(\widetilde{\ddot{\RV}} \mid \PHI) = \left(\frac{N^{M^2-M}}{(2 \pi)^{M^2-M} \det(\SIGMAM(\PHI))} \right)\frac{\exp\{-N[ \asin(\widetilde{\ddot{\RV}}) - \overline{\JM}\PSIM\arcsin(\PHI)]^H \SIGMAM^{-1}(\PHI) [ \asin(\widetilde{\ddot{\RV}}) - \overline{\JM}\PSIM\arcsin(\PHI)]\} }{ \Pi_{n=1}^{D-1} (1-[\PHI]_n^2)^{\nu_n} (1-[\PHI]_{n+D-1}^2)^{\nu_n} }.
\end{align}\normalsize
\vspace{-1mm}
\hrulefill
\vspace{-2mm}
\end{figure*}
Hence, the asymptotic ML estimation of $\PHI$ from $\widetilde{\ddot{\RV}}$ is derived as follows;
\small
\begin{align}
\label{Eq-Est-8}
   \widehat{\PHI} = ~ {\rm \underset{\PHI \in \mathds{E}_{\PHI}} {argmin}} ~L(\PHI),
\end{align}\normalsize
where the cost function $L(\PHI)$ is given in \eqref{Eq-Est-9} at the top of the next page in which $\nu_n = \|\VE (\LM_n)\|^2$.
\begin{figure*}[!t]
\vspace{-4mm}
\small
\begin{align}
\label{Eq-Est-9}
L(\PHI) = \ln \det(\SIGMAM(\PHI)) - \sum_{n=1}^{D-1} \nu_n \ln  (1-[\PHI]_n^2)(1-[\PHI]_{n+D-1}^2)+N[ \asin(\widetilde{\ddot{\RV}}) - \overline{\JM}\PSIM\arcsin(\PHI)]^H \SIGMAM^{-1}(\PHI) [ \asin(\widetilde{\ddot{\RV}}) - \overline{\JM}\PSIM\arcsin(\PHI)].
\end{align}\normalsize
\vspace{-1mm}
\hrulefill
\vspace{-2mm}
\end{figure*}
However, the minimization of \eqref{Eq-Est-9} with respect to $\PHI$ is very complicated owing to the nonlinearity of the cost function as well as the constraint $\PHI \in \mathds{E}_{\PHI}$. To make the problem computationally tractable, we first find an asymptotic equivalent approximation of $L(\PHI)$ which is much simpler to minimize. Let $\GAMMA \in \mathds{E}_{\GAMMA} \subset \mathds{R}^{(M^2-M) \times 1}$ be the $(M^2-M) \times 1$ vector containing  the real and imaginary parts of the elements of $\overline{\RM}$ above its main diagonal elements. Obviously, there is the following relationship between $\PHI$ and $\GAMMA$:
\small
\begin{align}
\label{Eq-Est-10}
    \GAMMA =  \FM \overline{\JM} \PSIM \PHI , \forall \PHI \in \mathds{E}_{\PHI},
\end{align}\normalsize
where $\FM = \frac{1}{2}\begin{bmatrix}
   \ddot{\FM}^T &
   \J \widetilde{\FM}^T
\end{bmatrix}^T \in \{0,1\}^{(M^2-M) \times (M^2-M)}
$
such that for all $1 \leq p < q \leq M $:
\begin{enumerate}
\item the \small$\left((p-1)M+q-\frac{p(p+1)}{2}\right)$\normalsize-th rows of $\ddot{\FM} \in \{0,1\}^{\frac{(M^2-M)}{2} \times (M^2-M)}$ is obtained by removing the elements with indices $(i-1)M+1$ for all $1 \leq i \leq M$ from $\overline{\EV}_p^T \otimes \overline{\EV}_q^T + \overline{\EV}_q^T \otimes \overline{\EV}_p^T$ with $[\overline{\EV}_p]_n = \delta[p-n]$ for $1 \leq n \leq M$.
\item the \small$\left((p-1)M+q-\frac{p(p+1)}{2}\right)$\normalsize-th rows of $\widetilde{\FM} \in \{0,1\}^{\frac{(M^2-M)}{2} \times (M^2-M)}$ is obtained by removing the elements with indices $(i-1)M+1$ for all $1 \leq i \leq M$ from $\overline{\EV}_p^T \otimes \overline{\EV}_q^T - \overline{\EV}_q^T \otimes \overline{\EV}_p^T$ with $[\overline{\EV}_p]_n = \delta[p-n]$ for $1 \leq n \leq M$.
\end{enumerate}
\begin{lem}
\label{lem-new-2}
The matrices $\FM$, $\PSIM$ and $\overline{\JM}$ are full rank.
\end{lem}
\begin{proof}
See Appendix \ref{app-new-1}.
\end{proof}
The mapping from $\PHI \in \mathds{E}_{\PHI}$ to $\GAMMA \in \mathds{E}_{\GAMMA}$ is one-to-one due to the full rankness of $\FM$, $\PSIM$ and $\overline{\JM}$. Hence, it is possible to equivalently reparameterize \eqref{Eq-Est-9} in terms of $\GAMMA$ instead of $\PHI$.  This can be done by simply replacing $\PHI$ with $\PSIM^{-1}\overline{\JM}^{\dagger}\FM^{-1}\GAMMA$. To achieve computational simplification, we make use of the fact that a consistent estimate of $\GAMMA$ can be obtained as $\widetilde{\GAMMA} = \FM \widetilde{\ddot{\RV}}$. We see that $\widetilde{\GAMMA} \in \mathds{R}^{(M^2-M) \times 1} \notin \mathds{E}_{\GAMMA}$ with probability one, since the $\mathds{E}_{\GAMMA}$ is a zero-measure
subset of $\mathds{R}^{(M^2-M) \times 1}$. Now,
considering the Taylor series expansion of $L(\GAMMA)$ around $\widetilde{\GAMMA}$, we obtain
\small
\begin{align}
 \label{Eq-Est-11}
L(\GAMMA) =& L(\widetilde{\GAMMA}) + (\GAMMA - \widetilde{\GAMMA})^H\nabla_{\GAMMA}L(\widetilde{\GAMMA}) \nonumber\\
&+\frac{1}{2}(\widetilde{\GAMMA} \!-\! \GAMMA)^H  \nabla_{\GAMMA}^2L(\widetilde{\GAMMA}) (\widetilde{\GAMMA} \!-\! \GAMMA) \!+\! \cdots\!,
\end{align}\normalsize
where $\nabla_{\GAMMA}L(\widetilde{\GAMMA})$ and $\nabla_{\GAMMA}^2L(\widetilde{\GAMMA})$ denote the gradient vector and the Hessian matrix of $L(\GAMMA)$ with respect to $\GAMMA$, computed at $\widetilde{\GAMMA}$, respectively. The first term in \eqref{Eq-Est-11} is constant and, moreover, the higher other term can be neglected for large $N$ considering the fact that $\widetilde{\GAMMA}$ is a consistent estimate of $\GAMMA$. Consequently, making use of \eqref{Eq-Est-10} and the fact that $\widetilde{\GAMMA} = \FM \widetilde{\ddot{\RV}}$, we have
\small
\begin{align}
 \label{Eq-Est-12}
\widehat{\PHI} \simeq ~& {\rm \underset{\PHI \in \mathds{E}_{\PHI}} {argmin}} ~( \overline{\JM} \PSIM \PHI - \widetilde{\ddot{\RV}})^H \FM^H \nabla_{\GAMMA}L(\widetilde{\GAMMA})  \nonumber\\
&+ (\widetilde{\ddot{\RV}} - \overline{\JM} \PSIM \PHI)^H \FM^H \nabla_{\GAMMA}^2L(\widetilde{\GAMMA})\FM (\widetilde{\ddot{\RV}} - \overline{\JM} \PSIM \PHI).
\end{align}\normalsize
The above quadratic optimization problem is asymptotically equivalent to \eqref{Eq-Est-8} but is much more convenient to work with. Relaxing the constraint $\PHI \in \mathds{E}_{\PHI}$ with $\PHI \in \mathds{R}^{2D-2}$ yields the following closed-form solution for $\widehat{\PHI}$
\small
\begin{align}
 \label{Eq-Est-13}
 \widehat{\PHI} \simeq & \PSIM^{-1}\left(\overline{\JM}^H \FM^H \nabla_{\GAMMA}^2L(\widetilde{\GAMMA}) \FM \overline{\JM} \right)^{-1} \overline{\JM}^H \nonumber\\
 & \times \left[\FM^H\nabla_{\GAMMA}^2L(\widetilde{\GAMMA})\FM \widetilde{\ddot{\RV}} - \FM^H\nabla_{\GAMMA}L(\widetilde{\GAMMA})\right].
\end{align}\normalsize
To derive the final expression for $\widehat{\PHI}$, we need to calculate $\nabla_{\GAMMA}L(\widetilde{\GAMMA})$ and $\nabla_{\GAMMA}^2L(\widetilde{\GAMMA})$. It is straightforward to derive $L(\GAMMA)$ by making use of \eqref{Eq-Est-10}. It follows that
\small
\begin{align}
 \label{Eq-Est-14}
  \nabla_{\GAMMA}L(\widetilde{\GAMMA}) =  \GV(\widetilde{\GAMMA}),
\end{align}\normalsize
where $[\GV(\GAMMA)]_n = \frac{4 [\GAMMA]_n}{1-|[\GAMMA]_n|^2} + \frac{\partial \ln \det(\SIGMAM(\GAMMA))}{\partial [\GAMMA]_n}$, for $1 \leq n \leq M^2-M$. Additionally, the Hessian matrix at $\widetilde{\GAMMA}$ is obtained as
\small
\begin{align}
 \label{Eq-Est-15}
\nabla_{\GAMMA}^2L(\widetilde{\GAMMA}) =& N   \DIAG(\widehat{\BV}) \FM^{-H} \widehat{\SIGMAM}^{-1} \FM^{-1} \DIAG(\widehat{\BV})  + \EM(\widetilde{\GAMMA}),
\end{align}\normalsize
where $\widehat{\SIGMAM} =\SIGMAM(\widetilde{\GAMMA})$,
$
[\widehat{\BV}]_n =\frac{1}{\sqrt{1-|[\widetilde{\GAMMA}]_n|^2}}
$, for $1 \leq n \leq M^2-M$ and $[\EM(\PHI)]_{n,l} = \frac{2\nu_n (1+|[\PHI]_n|^2)}{(1-|[\PHI]_n|^2)^2} + \frac{\partial^2 \ln \det(\SIGMAM(\GAMMA))}{\partial [\GAMMA]_n \partial [\GAMMA]_m}$.
Inserting \eqref{Eq-Est-14} and \eqref{Eq-Est-15} into \eqref{Eq-Est-13} leads to
\footnotesize
\begin{align}
 \label{Eq-Est-16}
&\widehat{\PHI} \!\simeq\! \PSIM^{-1} \left( \overline{\JM}^H \FM^H \DIAG(\widehat{\BV}) \FM^{-H} \widehat{\SIGMAM}^{-1} \FM^{-1} \DIAG(\widehat{\BV}) \FM \overline{\JM} \!+\!\overbrace{ \frac{ \EM(\widetilde{\GAMMA})}{N}}^{\hbar} \right)^{-1} \hspace{-4mm} \times \\
&\overline{\JM}^H \bigg(\FM^H \DIAG(\widehat{\BV}) \FM^{-H} \widehat{\SIGMAM}^{-1} \FM^{-1} \DIAG(\widehat{\BV}) \FM \widetilde{\ddot{\RV}}\!+\!\underbrace{\frac{\FM^H \EM(\widetilde{\GAMMA}) \FM \widetilde{\ddot{\RV}}- \FM^H\GV(\widetilde{\GAMMA})}{N}}_{\aleph}\bigg). \nonumber
\end{align}\normalsize
In the above equation, the terms $\hbar$ and $\aleph$ can be neglected for large $N$, thus \eqref{Eq-Est-16} may be simplified as
\small
\begin{align}
  \label{Eq-Est-17}
&\widehat{\PHI} \simeq \PSIM^{-1} \left( \overline{\JM}^H \FM^H \DIAG(\widehat{\BV}) \FM^{-H} \widehat{\SIGMAM}^{-1} \FM^{-1} \DIAG(\widehat{\BV}) \FM \overline{\JM} \right)^{-1}\nonumber\\
&\times \overline{\JM}^H \FM^H \DIAG(\widehat{\BV}) \FM^{-H} \widehat{\SIGMAM}^{-1} \FM^{-1} \DIAG(\widehat{\BV}) \FM \widetilde{\ddot{\RV}} .
\end{align}\normalsize
Hence, from \eqref{Eq-Est-5}, an enhanced consistent estimate of $\overline{\RV} = \VE(\overline{\RM})$ is derived as follows
\small
\begin{align}
\label{Eq-Est-18}
\widehat{\overline{\RV}} = \JM \begin{bmatrix} \ZEROV & \IDM_{D-1} & -\J \IDM_{D-1} \\
1 & \ZEROV & \ZEROV \\
\ZEROV & \IDM_{D-1} & \J \IDM_{D-1} \end{bmatrix} \begin{bmatrix} 1 \\ \widehat{\PHI}\end{bmatrix}.
\end{align}\normalsize
\begin{rmk}
\label{rmk-4}
Considering $\lim_{N \to \infty} \widetilde{\ddot{\RV}} = \ddot{\RV}$, it is readily observed from \eqref{Eq-Est-5} and \eqref{Eq-Est-17} that $\widehat{\PHI}$ is a consistent estimate of $\PHI$. This in turn implies that $\widehat{\overline{\RV}}$ is also a consistent estimate of $\overline{\RV}$.
\end{rmk}
To estimate DoAs using $\widehat{\overline{\RV}}$, we resort to CAB-MUSIC \cite{Liuonebit}. Specifically, we first construct the normalized augmented covariance matrix as
\small
\begin{align}
 \label{Eq-Est-19}
    \widehat{\overline{\RM}}_v = \begin{bmatrix}
   \mathbf{T}_v\mathbf{J}^{\dagger}\widehat{\overline{\RV}} & \mathbf{T}_{v-1}\mathbf{J}^{\dagger}\widehat{\overline{\RV}} & \cdots &
   \mathbf{T}_1\mathbf{J}^{\dagger}\widehat{\overline{\RV}}
\end{bmatrix} \in \mathds{C}^{v \times v},
\end{align}\normalsize
where $\mathbf{T}_i$ is a selection matrix, defined as
\small
\begin{align}\label{Eq-Est-20}
\hspace{-2mm}\mathbf{T}_i = \begin{bmatrix} \mathbf{0}_{v \times (i+D-v-1)} & \IDM_v & \mathbf{0}_{v \times (D-i)}\end{bmatrix} \in \{0,1\}^{v \times (2D-1)}. \hspace{-2mm}
\end{align}\normalsize
It follows from the consistency of $\widehat{\overline{\RV}}$ that
\small
\begin{align}
 \label{Eq-Est-21}
    \lim_{N \to \infty} \widehat{\overline{\RM}}_v &= \begin{bmatrix}
  \mathbf{T}_v\mathbf{J}^{\dagger}\overline{\RV} & \mathbf{T}_{v-1}\mathbf{J}^{\dagger}\overline{\RV} & \cdots &
  \mathbf{T}_1\mathbf{J}^{\dagger}\overline{\RV}
\end{bmatrix} \in \mathds{C}^{v \times v} \nonumber\\
&= \AM_v(\THETA) \DIAG(\overline{\PV}) \AM_v^H(\THETA) + \overline{\sigma}^2 \IDM_v,
\end{align}\normalsize
where $\AM_v(\THETA) = [\AV_v\left(\theta_1\right), \AV_v\left(\theta_2\right), \cdots, \AV_v\left(\theta_K\right)] \in \mathds{C}^{v \times K}$ denotes the steering matrix of a contiguous ULA with $v$ elements located at $(0 ,\frac{\lambda}{2}, \cdots, (v-1)\frac{\lambda}{2})$.
Hence, we can apply MUSIC to $\widehat{\overline{\RM}}_v$ to estimate the DoAs. We call the proposed method Enhanced One-bit CAB-MUSIC (EOCAB-MUSIC). Algorithm \ref{alg-1} summarizes the steps of EOCAB-MUSIC.
\begin{algorithm}[t]
\caption{EOCAB-MUSIC}
\begin{algorithmic}[1]
\qinput SLA one-bit observations, i.e., $\XM$.
\qoutput The estimates of source DoAs.
\State Compute the sample covariance matrix of one-bit data as $\widehat{\RM}_{\XV} = \frac{1}{N} \XM \XM^H$.
    \State Compute $\widetilde{\overline{\RM}}$ from \eqref{Eq-Est-3}.
    \State Form $\widetilde{\ddot{\RV}}$ by removing the diagonal elements of $\widetilde{\overline{\RM}}$ from $\VE(\widetilde{\overline{\RM}})$.
    \State Compute $\widetilde{\GAMMA}$ from $\widetilde{\GAMMA} = \FM \widetilde{\ddot{\RV}}$.
    \State Compute $\widehat{\BV}$ using $[\widehat{\BV}]_n =\frac{1}{\sqrt{1-|[\widetilde{\GAMMA}]_n|^2}}
$, for $1 \leq n \leq M^2-M$.
\State Compute $\widehat{\SIGMAM}$ by using (125) and replacing $\overline{\RM}$ with $\widetilde{\overline{\RM}}$ in (129), (130), (133) - (137), (139)-(148), (150) and (152)-(161) given in Appendix K.
\State Compute $\widehat{\PHI}$ from \eqref{Eq-Est-17}.
\State Compute $\widehat{\overline{\RV}}$ from \eqref{Eq-Est-18}.
\State Compute $\widehat{\overline{\RM}}_v$ from \eqref{Eq-Est-19}.
\State Apply MUSIC to $\widehat{\overline{\RM}}_v$ to estimate DoAs.
\end{algorithmic}
\label{alg-1}
\end{algorithm}
\begin{rmk}
The computational complexity of each step of Algorithm \ref{alg-1} is separately specified in Table \ref{table-1} where ${\cal G}(n)$, ${\cal K}(n)$ and ${\cal Z}$ denote the complexity of the chosen algorithm for multiplication of two $n$-digit numbers, the complexity of integration in (108) and the number of grid point of the MUSIC algorithm, respectively. Considering that $D$ and $v$ are typically in the order of $M^2$ and, moreover, $n$ and $M$ are normally very smaller than ${\cal Z}$, it follows from Table \ref{table-1} that the complexity of EOCAB-MUSIC is in the order of ${\cal O}( MN + M^2 ( {\cal G}(n) ( {\cal Z} + M^4) + {\cal K}(n) M^2 ) )$. On the other hand implementation of OCAB-MUSIC needs only steps 1, 2, 9 and 10 in algorithm \ref{alg-1}. Hence, its complexity is given by ${\cal O}( MN + M^2 {\cal G}(n) ( {\cal Z} + M^4) )$. Typically, we have ${\cal G}(n) ( {\cal Z} + M^4) \gg {\cal K}(n) M^2$, implying that the complexity of EOCAB-MUSIC is almost in the same order as that of OCAB-MUSIC.
\begin{table}[!]
\centering
\caption{Complexity of the steps of Algorithm \ref{alg-1}}
\begin{tabular}{ |c|c| }
\hline
Step order & Complexity\\
\hline
1 & ${\cal O}(MN)$\\
\hline
2 & ${\cal O}({\cal G}(n) \sqrt{n} M^2)$ \\
\hline
3 & ${\cal O}(M^2)$ \\
\hline
4 & ${\cal O}({\cal G}(n) M^4)$\\
\hline
5 & ${\cal O}({\cal G}(n) M^2)$\\
\hline
6 & ${\cal O}({\cal K}(n) M^4)$\\
\hline
7 & ${\cal O}({\cal G}(n) (DM^4+M^6+D^3) )$\\
\hline
8 & ${\cal O}({\cal G}(n) (D^2M^2)$\\
\hline
9 & ${\cal O}({\cal G}(n) (M^2v(2D-1+v))$\\
\hline
10 & ${\cal O}({\cal G}(n) ({\cal Z}M^2 + M^3) )$\\
\hline
\end{tabular}
\label{table-1}
\end{table}
\end{rmk}
\subsection{Asymptotic Performance Analysis}
\label{sec:MSE}
In this section, we investigate the asymptotic performance of the proposed estimator through the derivation of a closed-form expression for the second-order statistics of the asymptotic distribution (as $N \to \infty$) of the DoA estimation errors. Our main results are summarized in Theorem \ref{Theo-6}, Corollary \ref{Col-1} and Theorem \ref{Theo-7}.
\begin{lem}
\label{lem-2}
$\widehat{\THETA}$ obtained by EOCAB-MUSIC is a consistent estimate of $\THETA$ if $K \leq v- 1$.
\end{lem}
 \begin{proof}
See Appendix \ref{App-F}
\end{proof}
\begin{theo}
\label{Theo-6}
The closed-form expression for the covariance of the asymptotic distribution (as $N \to \infty$) of the DoA estimation errors obtained by EOCAB-MUSIC is given by
\small
\begin{align}
\label{Eq-cov}
&{\cal E}_{\theta_{k_1}, \theta_{k_2}} = \EX\{(\theta_{k_1} - \widehat{\theta}_{k_1})(\theta_{k_2} - \widehat{\theta}_{k_2})^*\} \\
&= \frac{ (\sigma^2+\sum_{k=1}^K p_k)^2}{ N \pi^2 p_{k_1} p_{k_2} q_{k_1} q_{k_2} \cos \theta_{k_1} \cos \theta_{k_2} }\nonumber\\
&\times \Re\{\ZV^T_{k_1} \overline{\TM} (\overline{\JM}^H \WM \overline{\JM})^{-1} \overline{\JM}^H \WM \GAMMAM \WM \overline{\JM} (\overline{\JM}^H \WM \overline{\JM})^{-1} \overline{\TM}^H \ZV_{k_2}^* \},\nonumber
\end{align}\normalsize
where
\small
\begin{align}
\ZV_k =& \BETA_k \otimes \ALPHA_k,\\
\BETA_k =& \Pi^{\perp}_{\scaleto{\AM_v(\THETA)\mathstrut}{5pt}} \DIAG(\VV) \AV_v(\theta_k),\\
\ALPHA_k =&  \AM_v^{\dagger T}(\THETA) \IMATH_k,
\end{align}
\begin{align}
q_k =& \AV_v^H(\theta_k) \DIAG(\VV) \Pi^{\perp}_{\AM_v} \DIAG(\VV) \AV_v(\theta_k),\\
\label{Eq-Est-22}
\WM =& \FM^H \DIAG(\BV) \FM^{-H} \SIGMAM^{-1} \FM^{-1} \DIAG(\BV) \FM,\\
\label{Eq-Est-23}
[\GAMMAM]_{p,q} =& \frac{1}{2}  \bigg(\sqrt{1-[\Re\{[\ddot{\RV}]_p\}]^2} \times \sqrt{1-[\Re\{[\ddot{\RV}]_q\}]^2}\\
&+ \sqrt{1-[\Im\{[\ddot{\RV}]_p\}]^2} \times \sqrt{1-[\Im\{[\ddot{\RV}]_q\}]^2}\bigg) \Re\{[\SIGMAM]_{p,q}\} \nonumber\\
&+\frac{\J}{2} \bigg(\sqrt{1-[\Im\{[\ddot{\RV}]_p\}]^2} \times \sqrt{1-[\Re\{[\ddot{\RV}]_q\}]^2} \nonumber\\
&+ \sqrt{1-[\Re\{[\ddot{\RV}]_p\}]^2} \times \sqrt{1-[\Im\{[\ddot{\RV}]_q\}]^2}\bigg) \Im\{[\SIGMAM]_{p,q}\}, \nonumber
\end{align}\normalsize
with $\VV = [0,1,2, \cdots, v-1]^T$, $
[\BV]_n =\frac{1}{\sqrt{1-|[\GAMMA]_n|^2}}
$ for $1 \leq n \leq M^2-M$, $\SIGMAM \in \mathds{C}^{(M^2-M) \times (M^2-M)}$ as given in Appendix K (kindly refer to the supplementary document), $\overline{\TM} \in \mathds{C}^{v^2 \times (2D-2)}$ as defined in \eqref{Eq-app-G-9-1} in Appendix \ref{app-G}, and $\IMATH_k$ being the $k^{\rm th}$ column of $\IDM_K$.
\end{theo}
 \begin{proof}
See Appendix \ref{app-G}
\end{proof}
\begin{Col}
\label{Col-1}
The asymptotic MSE expression (as $N \to \infty$) for the DoA estimates obtained by EOCAB-MUSIC is given by
\small
\begin{align}
\label{Eq-mse}
{\cal E}_{\theta_k} &= \EX\{(\theta_{k_1} - \widehat{\theta}_k)^2\} = \frac{(\sigma^2+\sum_{k'=1}^K p_{k'})^2}{N \pi^2 p_k^2 q_k^2 \cos^2 \theta_k} \\
&\times \Re\{\ZV^T_k \overline{\TM} (\overline{\JM}^H \WM \overline{\JM})^{-1} \overline{\JM}^H \WM \GAMMAM \WM \overline{\JM} (\overline{\JM}^H \WM \overline{\JM})^{-1} \overline{\TM}^H \ZV_k^* \}.\nonumber
\end{align}\normalsize
\end{Col}
\begin{Col}
\label{Col-2}
The covariance of the asymptotic distribution (as $N \to \infty$) of the DoA estimation errors and the asymptotic MSE expression (as $N \to \infty$) for the one-bit DoA estimator given in \cite{Liuonebit}, named as One-bit CAB-MUSIC (OCAB-MUSIC), is easily obtained by replacing $\WM$ with $\IDM_{M^2-M}$ in \eqref{Eq-cov} and \eqref{Eq-mse}, respectively.
\end{Col}
 \begin{proof}
See Appendix \ref{app-I}
\end{proof}
\begin{rmk}
\label{rmk-5}
It is concluded from Corollary \ref{Col-1} and Corollary \ref{Col-2} that, similar to Infinite-bit Co-Array-Based MUSIC (ICAB-MUSIC) \cite{Wang2017}, the MSEs of EOCAB-MUSIC and OCAB-MUSIC depend on both the physical and the virtual array geometries through $\AM_v(\theta)$ and $\overline{\RM}$, respectively.
\end{rmk}
\begin{rmk}
\label{rmk-6}
Another interesting implication of Corollary \ref{Col-1} is that the MSEs of EOCAB-MUSIC and OCAB-MUSIC reduce at the same rate as that of ICAB-MUSIC \cite{Wang2017} with respect to $N$; i.e. ${\cal E}_{\theta_k} \propto \frac{1}{N}$ for both.
\end{rmk}
\begin{rmk}
\label{rmk-7}
It is readily clear from the definition that $\overline{\RV}$ is a function of the SNR, and not $\PV$ and $\sigma^2$. This indicates that $\WM$ and $\GAMMAM$ are also functions of the SNR instead of $\PV$ and $\sigma^2$. Further, multiplying the numerator and denominator of $(\sigma^2\!+\!\sum_{k'=1}^K p_{k'})^2/p_k^2$ by $1/\sigma^4$ reformulates it as a function of the SNR. These observations imply that the MSEs of EOCAB-MUSIC and OCAB-MUSIC are functions of the SNR instead of $\PV$ and $\sigma^2$. This fact can also be deduced directly from system model where we have
\small
\begin{align}
  \hspace{-1mm}[\XV(t)]_m \!=\! \frac{1}{\sqrt{2}} \SGN\left(\Re\{[\YV(t)]_m\} \right)
    \!+\! \frac{\J}{\sqrt{2}}~ \SGN\left(\Im\{[\YV(t)]_m \}\right)\nonumber\\
    = \frac{1}{\sqrt{2}} \SGN\left(\Re\{\frac{[\YV(t)]_m}{\sigma}\} \right)
    \!+\! \frac{\J}{\sqrt{2}}~ \SGN\left(\Im\{\frac{[\YV(t)]_m}{\sigma} \}\right).
\end{align}\normalsize
for $\sigma > 0$. This implies that, without loss of generality, we can consider the power of each source equal to the SNR for that source and the noise variance equal to $1$.
\end{rmk}
\begin{theo}
\label{Theo-7}
Assume all sources have equal power $p$ and $SNR=p/\sigma^2$. Then, for a sufficiently large SNR, the MSE of EOCAB-MUSIC converges to the following constant value:
\small
\begin{align}
&\lim_{SNR \to \infty} {\cal E}_{\theta_k} = \frac{K^2}{N \pi^2 q_k^2 \cos^2 \theta_k} \times \\
& \Re\{\ZV^T_k \overline{\TM} (\overline{\JM}^H \WM_{\infty} \overline{\JM})^{-1} \overline{\JM}^H \WM_{\infty} \GAMMAM_{\infty} \WM_{\infty} \overline{\JM} (\overline{\JM}^H \WM_{\infty} \overline{\JM})^{-1} \overline{\TM}^H \ZV_k^* \}\!>\!0,\nonumber
\end{align}\normalsize
where $\WM_{\infty}$ and $\GAMMAM_{\infty}$ are obtained by replacing $\overline{\RM}$, $\ddot{\RV}$ and $\GAMMA$ in the definitions of $\WM$ and $\GAMMAM$ (kindly refer to Theorem \ref{Theo-6}) with $\overline{\RM}_{\infty}$, $\GAMMA_{\infty}$ and $\ddot{\RV}_{\infty}$, respectively, where
\small
\begin{align}
\overline{\RM}_{\infty} = \frac{1}{K} \AM(\THETA) \AM^H(\THETA) + (1-\frac{1}{K}) \IDM_M,
\end{align}\normalsize
$\GAMMA_{\infty}$ is the $(M^2-M) \times 1$ vector containing  the real and imaginary parts of the elements of $\overline{\RM}_{\infty}$ above its main diagonal elements and $\ddot{\RV}_{\infty} = \PSIM^{-1}\overline{\JM}^{\dagger}\FM^{-1} \GAMMA_{\infty}$.
\end{theo}
 \begin{proof}
See Appendix \ref{app-J}.
\end{proof}
\begin{rmk}
\label{rmk-8}
It follows from Theorem \ref{Theo-7} that it is not possible to make the MSEs of EOCAB-MUSIC and OCAB-MUSIC arbitrarily small by increasing the SNR.
\end{rmk}
\section{Simulation Results}
\label{sec:simulations}
In this section, we provide some numerical results to validate the analytical results obtained in previous sections as well as to assess the performance of the proposed DoA estimator. Specifically, we will show that the proposed estimator yields better performance in terms of estimation accuracy and resolution compared to the approach given in \cite{Liuonebit}. In the rest of this section, we will refer to: \begin{enumerate*} \item the CRB for DoA estimation from infinite-bit measurements as Infinite-bit CRB (I-CRB), whose expression is given in Remark \ref{rmk-new-new-2}; \item the pessimistic approximation of the CRB for DoA estimation from one-bit measurements as One-bit CRB (O-CRB);
\item CAB-MUSIC using infinite-bit measurements as Infinite-bit CAB-MUSIC (ICAB-MUSIC); \item the DoA estimator given in \cite{Liuonebit} as one-bit CAB-MUSIC (OCAB-MUSIC); \item the proposed estimator in this paper as Enhanced One-bit CAB-MUSIC (EOCAB-MUSIC) \end{enumerate*}.
\subsection{General Set-up}
In all experiments, each simulated point has been computed by $5000$ Monte Carlo repetitions. Unless the source locations are specified for a particular result, it is assumed that the $K$ independent sources are equally spaced in the angular domain $[-\ang{60}, \ang{60}]$ such that $\theta = -\ang{60}$ when $K=1$. Further, all sources are assumed to have equal powers, i.e., $p_k = p$ for all $k$, and the SNR is defined as $10 \log  \frac{p}{\sigma^2}$. For our numerical investigation, we use four different types of arrays with $M=10$ physical elements and the following geometries:
\small
\begin{align}
\label{nested}
&\mathds{M}_{\text{nested}}: \left\{1, 2, 3, 4, 5, 6, 12, 18, 24, 30\right\}, \\
\label{co-prime}
&\mathds{M}_{\text{co-prime}}: \left\{0, 3, 5, 6, 9, 10, 12, 15, 20, 25\right\}, \\
\label{MRA}
&\mathds{M}_{\text{MRA}}: \left\{0, 1, 3, 6, 13, 20, 27, 31, 35, 36 \right\},\\
\label{ULA}
&\mathds{M}_{\text{ULA}}: \left\{0, 1, 2, \cdots, 9\right\}.
\end{align}\normalsize
These arrays generate the difference co-arrays:
\small
\begin{align}
\label{co-nested}
&\mathds{D}_{\text{nested}}: \left\{0,1,2, \cdots, 29\right\}, \\
\label{co-co-prime}
&\mathds{D}_{\text{co-prime}}: \left\{0,1, 2, \cdots, 22, 25 \right\}, \\
\label{co-MRA}
&\mathds{D}_{\text{MRA}}: \left\{0,1, 2, \cdots, 36 \right\},
\\
\label{co-ULA}
&\mathds{D}_{\text{ULA}}: \left\{0,1, 2, \cdots, 9 \right\}.
\end{align}\normalsize
Further, we generate the grid from $-\ang{90}$ to $\ang{90}$ with step size $\ang{0.001}$ to implement MUSIC.
To avoid griding, alternatively, it is also possible to use root-MUSIC.
\begin{figure*}[!]
\centering
\subfloat[]{\includegraphics[width=0.7\columnwidth]{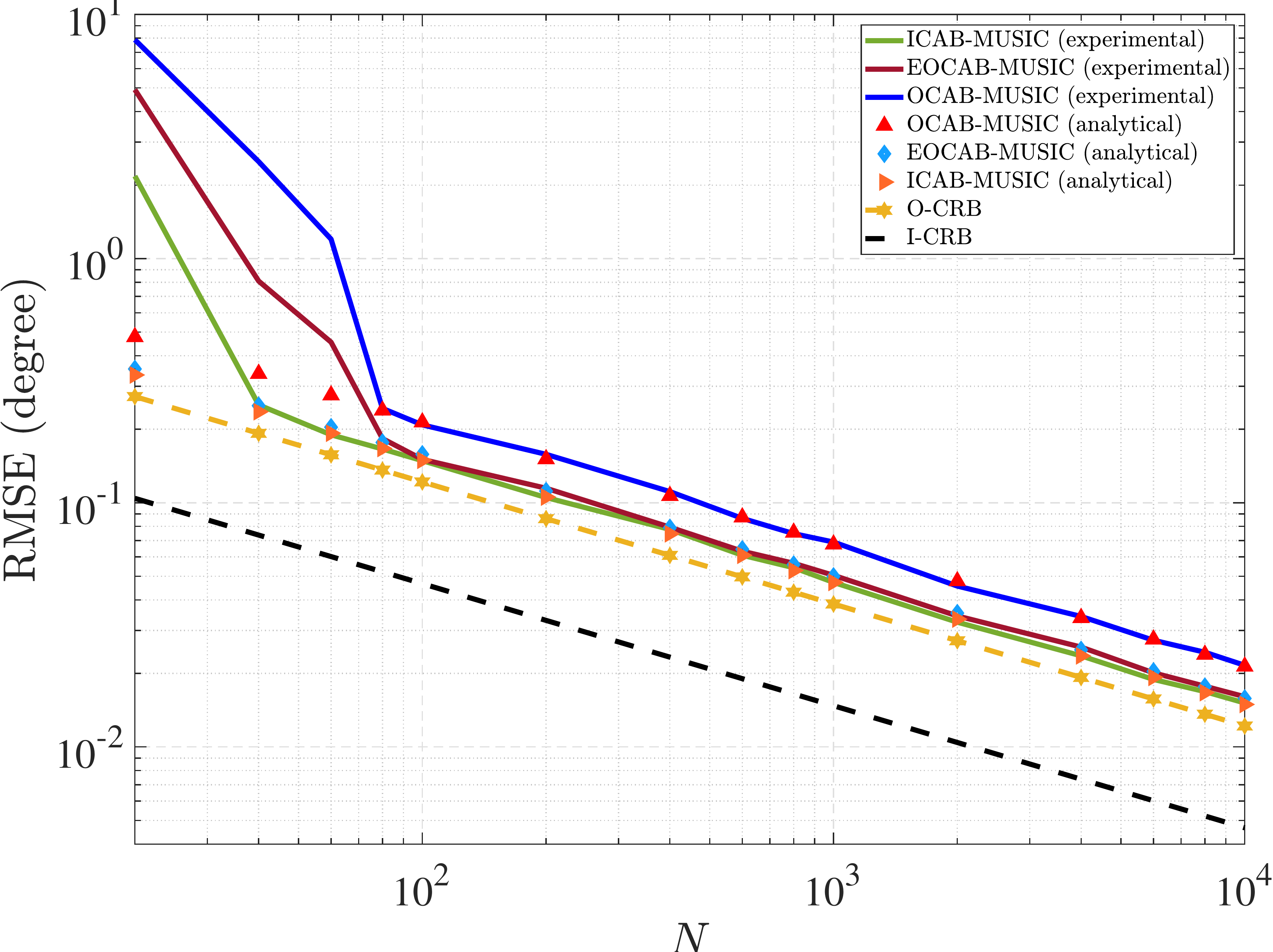}%
\label{Fig:2-a}}
\hfil
\subfloat[]{\includegraphics[width=0.7\columnwidth]{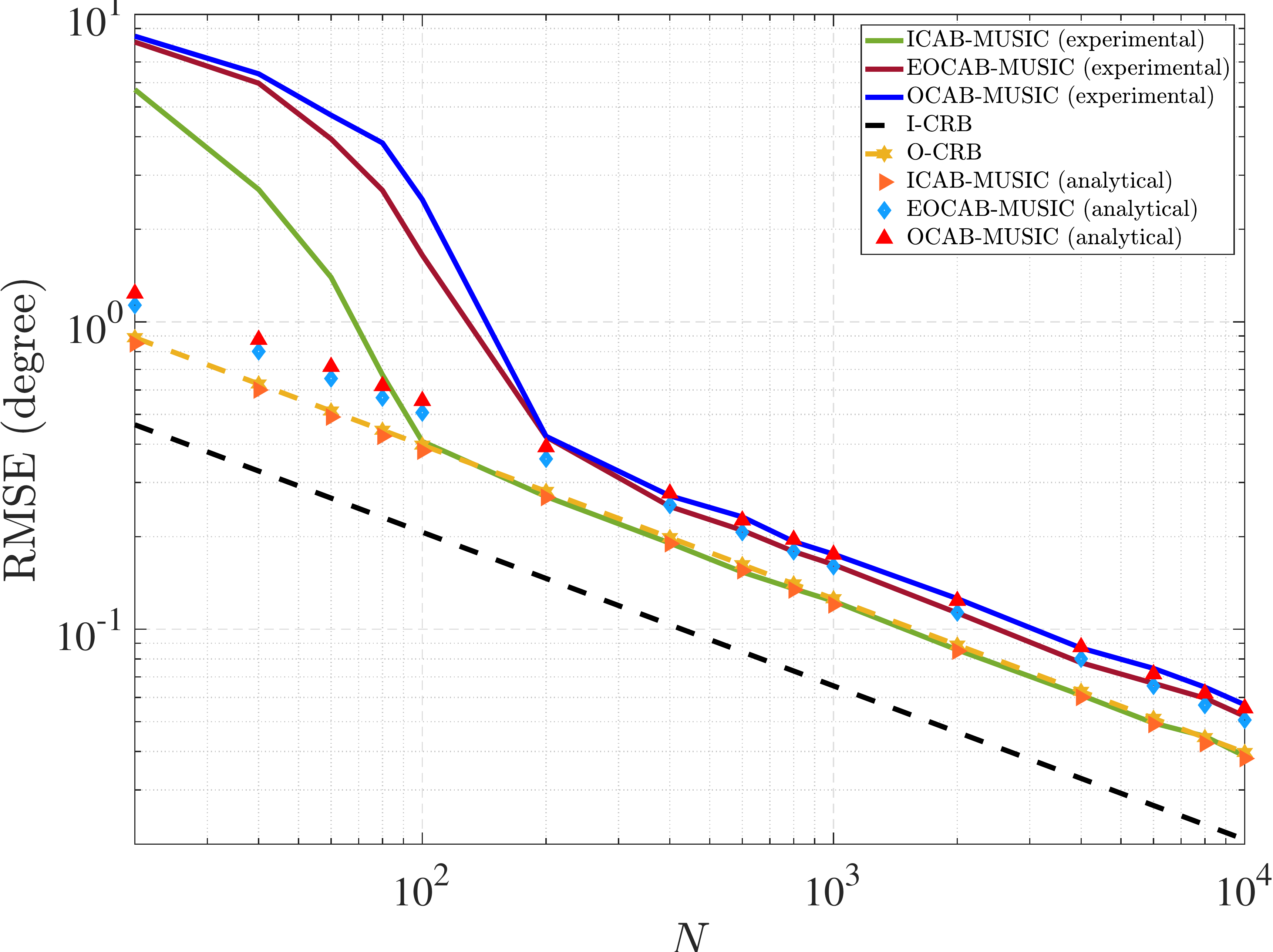}%
\label{Fig:2-b}}
\caption{RMSE in degrees for $\theta_2$ versus $N$ for a nested array with $M=10$ elements and configuration given in (\ref{nested}), ${\rm SNR}= 3$ dB, and: (a) $K=5<M$; (b) $K=12>M$.}
\label{Fig:2}
\vspace{-4mm}
\end{figure*}
\begin{figure}[!]
\centering
\includegraphics[width=0.7\columnwidth]{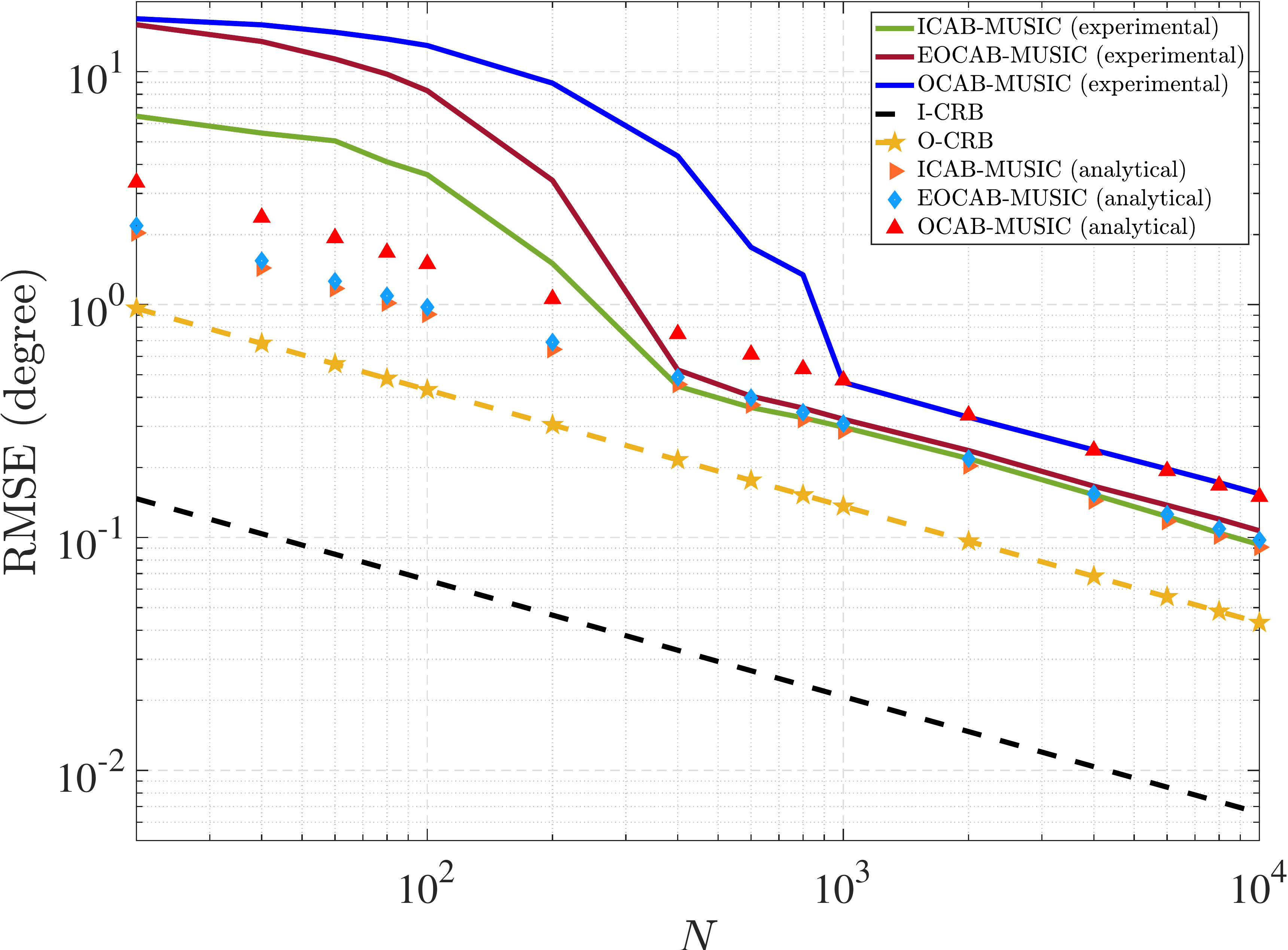}
\DeclareGraphicsExtensions.
\caption{RMSE in degrees for $\theta_2$ versus $N$ for a nested array with $M=10$ elements and configuration given in (\ref{nested}) when $K=3$, $\theta_1=\ang{2}$, $\theta_2=\ang{3}$, $\theta_3=\ang{75}$, ${\rm SNR}_1= 20$ dB, ${\rm SNR}_2= 8$ dB and ${\rm SNR}_3= 22$ dB.}
\label{Fig:8}
\end{figure}
\subsection{MSE vs. the Number of Snapshots}
Fig. \ref{Fig:2} depicts the Root-Mean-Squares-Error (RMSE) for $\theta_2$ in degree versus the number of snapshots when the nested array in (\ref{nested}) is used. The SNR is assumed to be $3$ dB. In addition, noting $M=10$, two different scenarios are considered: (a) $K=5<M$, and (b) $K=12>M$. Fig. \ref{Fig:2} illustrates a close agreement between the numerical simulations and analytical expression derived for RMSEs of OCAB-MUSIC and EOCAB-MUSIC when about $200$ or more snapshots are available. Further, a considerable gap is observed between the performance of OCAB-MUSIC and that of the EOCAB-MUSIC. For instance, at $N=400$, Figs. \ref{Fig:2-a} and \ref{Fig:2-b} show a performance gain of roughly $3$ dB and $1$ dB, respectively, in terms of the RMSE when the EOCAB-MUSIC is used. It is also observed that EOCAB-MUSIC performs as well as ICAB-MUSIC when $K=5<M$. Further, it is observed that the RMSE of EOCAB-MUSIC is very close to O-CRB when $K=5<M$ but we see a gap between them when $K=12>M$.

Fig. \ref{Fig:2} also shows that when a small number of snapshots is available, e.g. less than $1000$, all estimators are confronted with substantial performance degradation. This performance loss is justified by the subspace swap arising from the inaccurate estimate of the normalized covariance matrix of $\YV(t)$, i.e. $\overline{\RM}$, in this case. However, it is seen that the proposed estimator still has superior performance compared to OCAB-MUSIC, even in the low snapshot paradigm.

Fig. \ref{Fig:8} depicts the RMSE $\theta_2$ in degree versus the number of snapshots when $K=3$ and the sources powers are unequal. Specifically, It is assumed that $\theta_1=\ang{2}$, $\theta_2=\ang{3}$, $\theta_3=\ang{75}$, ${\rm SNR}_1= 20$ dB, ${\rm SNR}_2= 8$ dB and ${\rm SNR}_3= 22$ dB. Comparing Fig. \ref{Fig:2} with Fig. \ref{Fig:8} reveals that a high difference between the SNRs of the closely-spaced source signals do not have a meaningful impact on the relative asymptotic performance of ICAB-MUSIC, OCAB-MUSIC and EOCAB-MUSIC, however, by increasing the difference between SNRs, OCAB-MUSIC needs more number of snapshots to achieve its asymptotic performance compared to EOCAB-MUSIC and ICAB-MUSIC.
%
\subsection{MSE vs. SNR}
\begin{figure*}[!]
\centering
\subfloat[]{\includegraphics[width=0.7\columnwidth]{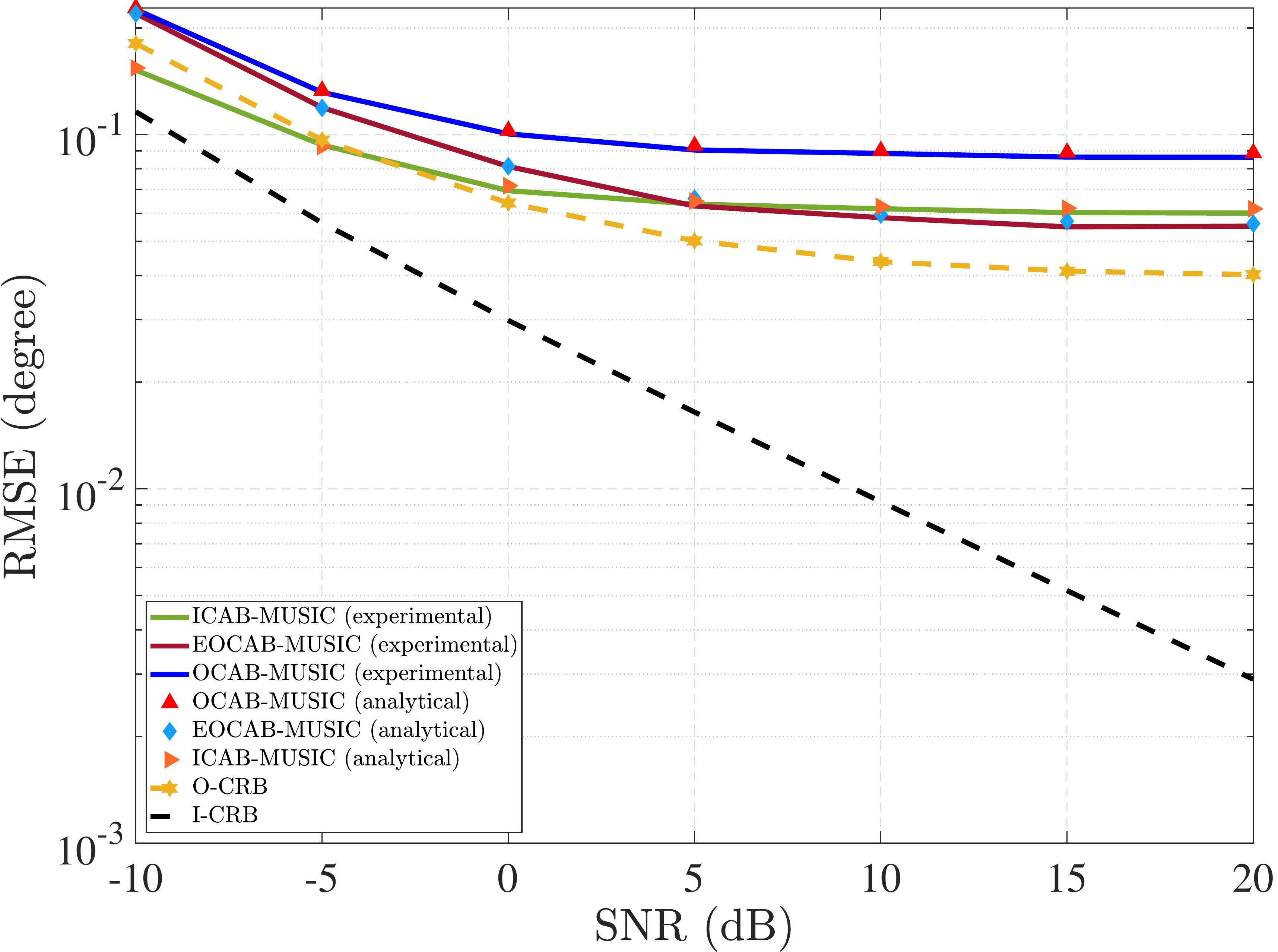}%
\label{Fig:3-a}}
\hfil
\subfloat[]{\includegraphics[width=0.7\columnwidth]{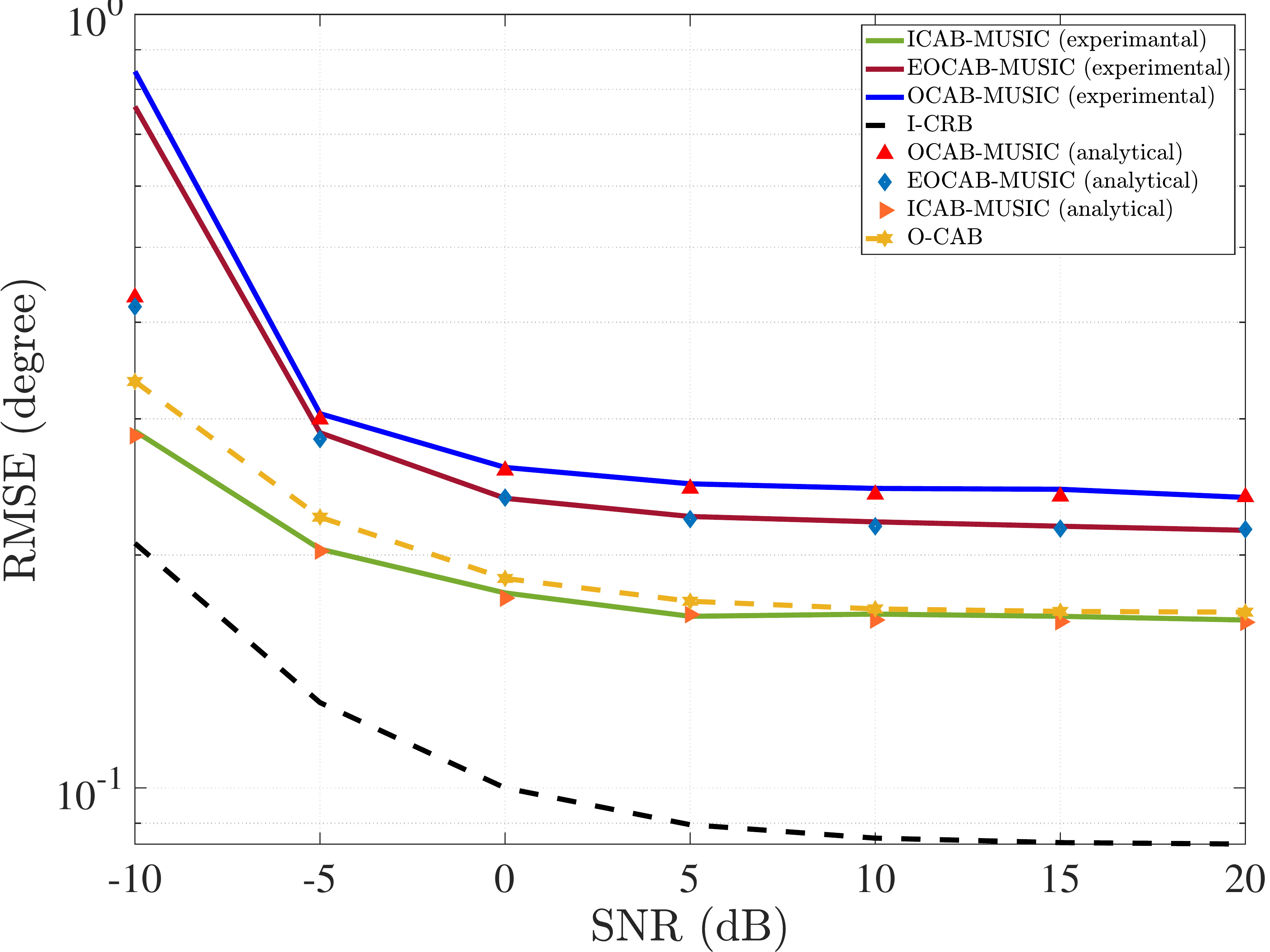}%
\label{Fig:3-b}}
\caption{RMSE in degrees for $\theta_2$ versus SNR wen the source powers are equal for a nested array with $M=10$ elements and configuration given in (\ref{nested}), $N=500$, and: (a) $K=5<M$; (b) $K=12>M$.}
\label{Fig:3}
\vspace{-3mm}
\end{figure*}

Fig. \ref{Fig:3} shows the RMSE for $\theta_2$ in degrees versus SNR for the same setup used for Fig. \ref{Fig:2}. The number of snapshots is considered to be $N=500$. It is seen in Figs. \ref{Fig:3-a} and Fig. \ref{Fig:3-b} that the RMSEs of OCAB-MUSIC and EOCAB-MUSIC perfectly match with their asymptotic analytical RMSEs given in Corollary \ref{Col-1} and Corollary \ref{Col-2}.

Fig. \ref{Fig:3} demonstrates that the I-CRB tends to decay to zero as the SNR increases when $K=5<M$ while it gets saturated as the SNR increases when $K=12>M$. However, as opposed to the I-CRB, O-CRB tends to converge to a constant non-zero value at the high SNR regime for both the cases $K=5<M$ and $K=12>M$. This behavior of O-CRB was already predicted by Theorem \ref{Theo-5}. In addition, as shown in Theorem \ref{Theo-7}, the RMSEs of OCAB-MUSIC and EOCAB-MUSIC also converge to a constant non-zero value as the SNR  increases for both $K=5<M$ and $K=12>M$.

We observe from Fig. \ref{Fig:3} that EOCAB-MUSIC preforms better than OCAB-MUSIC in both scenarios $K=5<M$ and $K=12>M$. For example, at ${\rm SNR}=5$, EOCAB-MUSIC leads to performance gains of about $3.7$ dB and $1.15$ dB in terms of RMSE compared to OCAB-MUSIC. Further, it is seen that EOCAB-MUSIC even outperforms ICAB-MUSIC at high SNR regime when $K=5<M$. Another interesting observation is that the RMSE of O-CRB is either better or equal to that of ICAB-MUSIC.

\begin{figure*}[!]
\centering
\subfloat[]{\includegraphics[width=0.7\columnwidth]{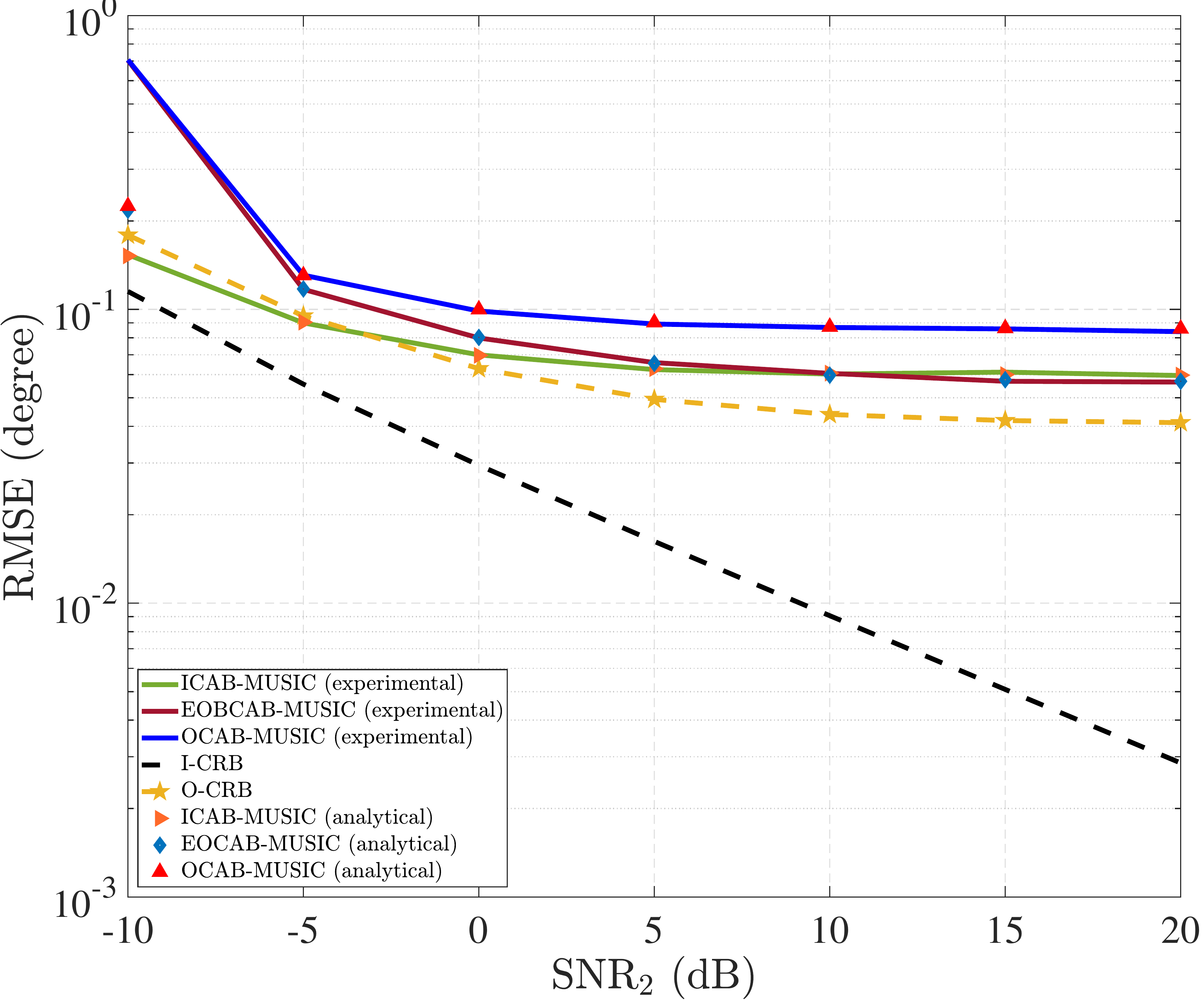}%
\label{Fig:6-a}}
\hfil
\subfloat[]{\includegraphics[width=0.7\columnwidth]{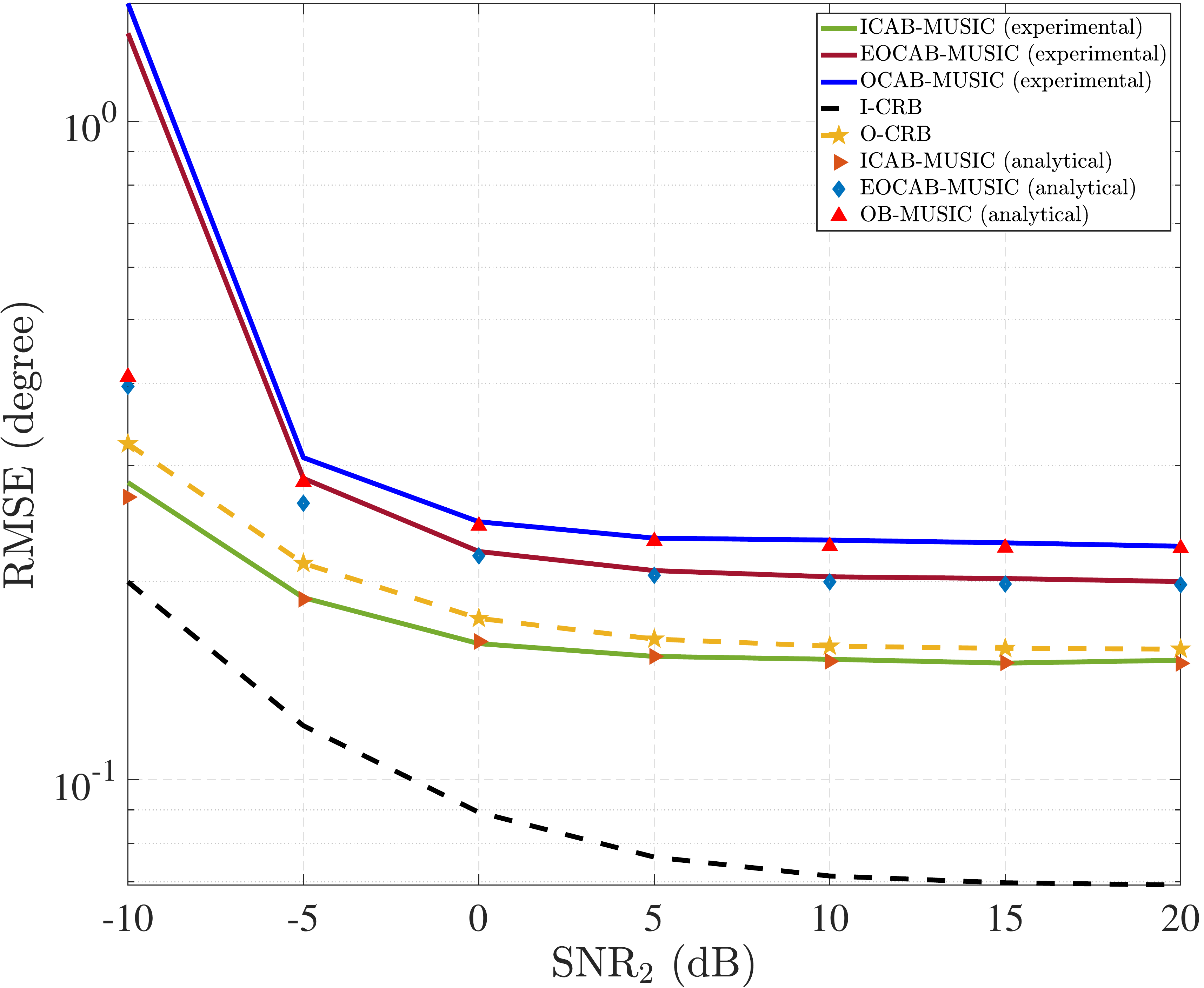}%
\label{Fig:6-b}}
\caption{RMSE in degrees for $\theta_2$ versus SNR when the source powers are unequal for a nested array with $M=10$ elements and configuration given in (\ref{nested}), $N=500$, and: (a) $K=5<M$; (b) $K=12>M$.}
\label{Fig:6}
\vspace{-3mm}
\end{figure*}

Fig. \ref{Fig:6} shows the RMSE for $\theta_2$ in degrees versus SNR when the sources powers are unequal and DoAs are not exactly on the grid as opposed to Fig. \ref{Fig:3}. The number of snapshots is considered to be $N=500$. In case of $K=5<M$, the sources are located at $\theta_1= \ang{-49.4551}, \theta_2=\ang{-30.1443}, \theta_3=\ang{-2.4525}, \theta_4=\ang{26.8293}$ and $\theta_5=\ang{56.5149}$. Further, the source SNRs are assumed to be ${\rm SNR}_1= 0.75 \times {\rm SNR}_2, {\rm SNR}_3=1.22 \times {\rm SNR}_2, {\rm SNR}_4= 0.92 \times {\rm SNR}_2$ and ${\rm SNR}_5= 0.66 \times {\rm SNR}_2$ while ${\rm SNR}_2$ varies from $10$ dB to $20$ dB as shown in Fig. \ref{Fig:6-a}. Further, in case of $K=12>M$, the sources are located at $\theta_1= \ang{-56.3351}, \theta_2=\ang{-36.2628}, \theta_3=\ang{-19.9004}, \theta_4=\ang{-2.4093}, \theta_5=\ang{0.0027}, \theta_6=\ang{13.1840}, \theta_7=\ang{23.8495}, \theta_8=\ang{25.8044}, \theta_9=\ang{29.2889}, \theta_10=\ang{40.9107},
\theta_11=\ang{48.4465}$ and $\theta_12=\ang{48.5667}$. The source SNRs are assumed to be ${\rm SNR}_1= 1.34 \times {\rm SNR}_2, {\rm SNR}_3=0.84 \times {\rm SNR}_2, {\rm SNR}_4= 0.83 \times {\rm SNR}_2, {\rm SNR}_5= 0.67 \times {\rm SNR}_2, {\rm SNR}_6= 0.69 \times {\rm SNR}_2, {\rm SNR}_7= 0.95 \times {\rm SNR}_2, {\rm SNR}_8= 0.61 \times {\rm SNR}_2, {\rm SNR}_9= 0.79 \times {\rm SNR}_2, {\rm SNR}_10= 0.56 \times {\rm SNR}_2, {\rm SNR}_10= 0.82 \times {\rm SNR}_2$ and ${\rm SNR}_12= 0.88 \times {\rm SNR}_2$ while ${\rm SNR}_2$ varies from $10$ dB to $20$ dB as shown in Fig. \ref{Fig:6-b}. Comparing Fig. \ref{Fig:6} with Fig. \ref{Fig:3} reveals that unequal source powers do not have remarkable impact on the estimation accuracy particularly in high-SNR regime.
%
\subsection{CRB vs. the Number of Source Signals}
\begin{figure}[!]
\centering
\includegraphics[width=0.7\columnwidth]{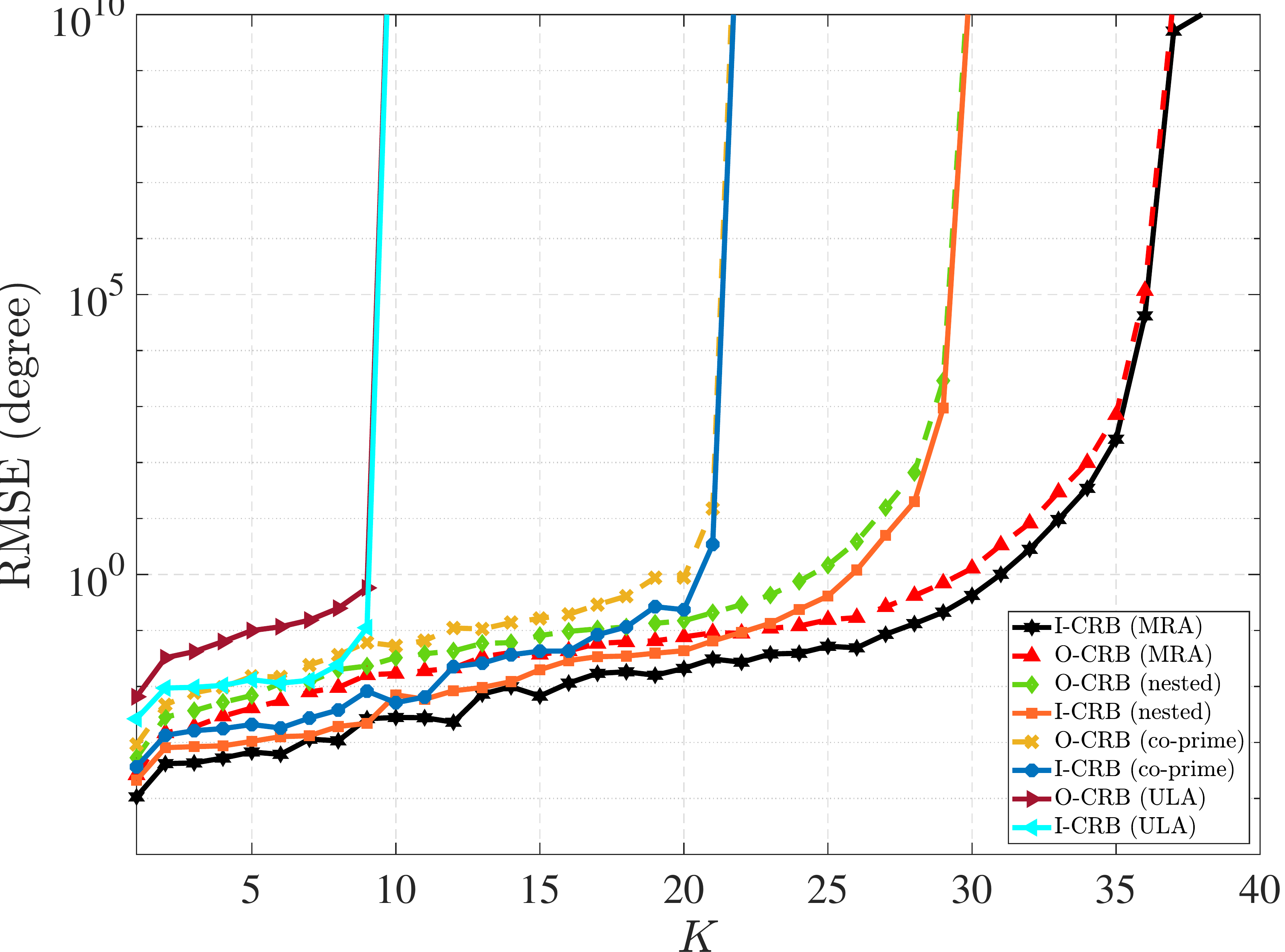}
\DeclareGraphicsExtensions.
\caption{The CRB versus $K$ for various array configurations given from \eqref{nested} to \eqref{MRA}, $N=500$ and  ${\rm SNR}= 3$ dB.}
\label{Fig:4}
\end{figure}

Fig. \ref{Fig:4} plots the I-CRB and the O-CRB for $\theta_2$ in degree versus the number of source signals for ${\rm SNR} = 3~{\rm dB}$ and $N=500$ and different types of arrays given in \eqref{nested}, \eqref{co-prime}, \eqref{MRA} and \eqref{ULA}. The values of $D$ and $v$ for the different types of arrays are as: \begin{enumerate*}
\item MRA: $D=37$ and $v=37$; \item nested array: $D=30$ and $v=30$; \item co-prime array: $D=26$ and $v=23$; ULA: $D=10$ and $v=10$ \end{enumerate*}. Fig. \ref{Fig:4} indicates that both the I-CRB and the O-CRB increase as the number of source signals increases. Moreover, it is observed that the I-CRB and the O-CRB are quite small for all the SLAs as long as $1 \leq K \leq v-1$, but they escalate dramatically when $K$ approaches values that are equal to or larger than $D$. This observation is in compliance with Theorem \ref{Theo-2} which indicates that the DoA estimation problem is globally identifiable when $1\leq K \leq v-1$ and is globally non-identifiable when $K \geq D$.
\subsection{Resolution Probability}
\begin{figure}[!]
\centering
\includegraphics[width=0.7\columnwidth]{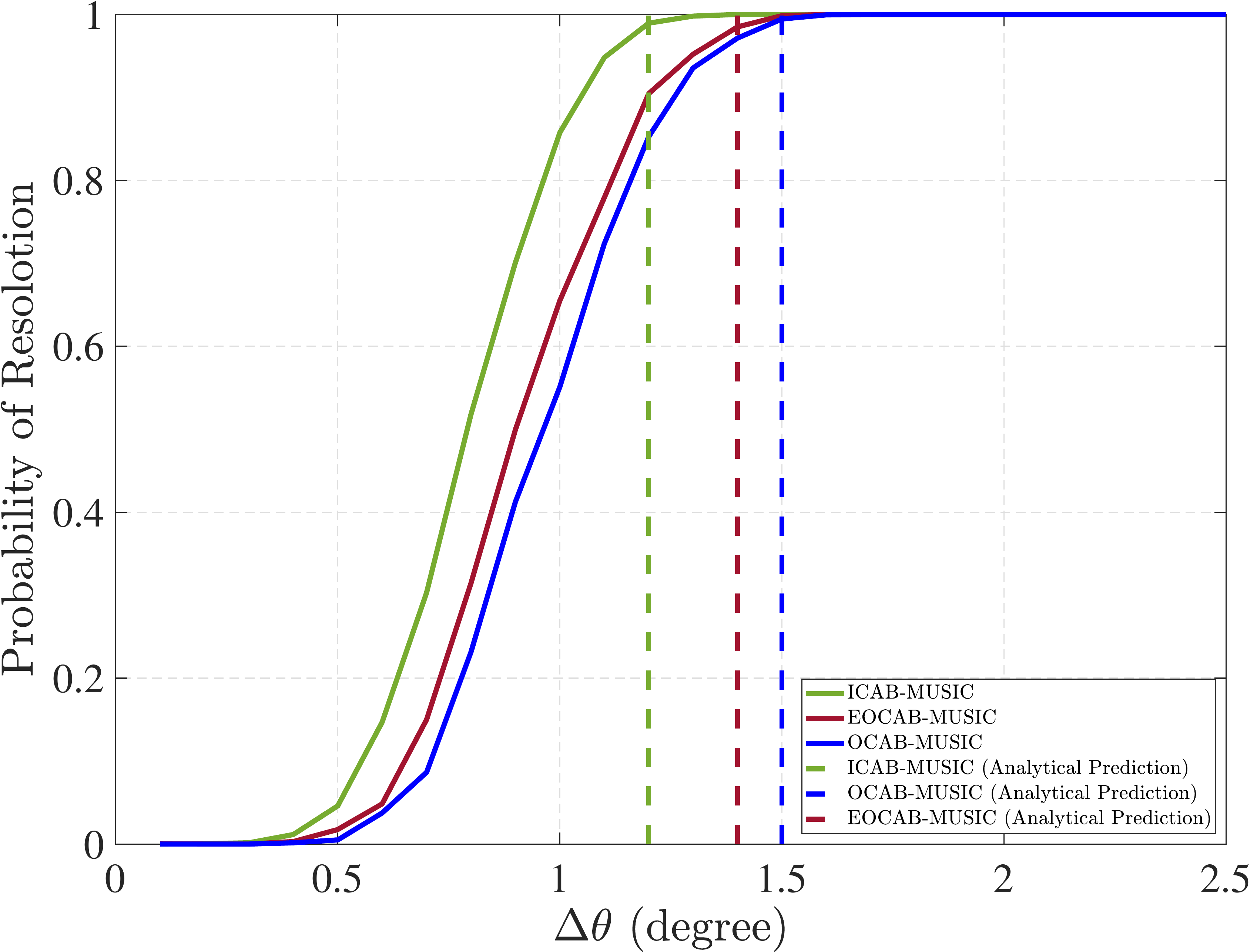}
\DeclareGraphicsExtensions.
\caption{Probability of resolution versus source separation in degree for a nested array with $M=10$ elements and configuration given in (\ref{nested}), $N=500$ and  ${\rm SNR}= 0$ dB.}
\label{Fig:5}
\vspace{-2mm}
\end{figure}

Fig. \ref{Fig:5} depicts the probability of resolution versus the source separation for ICAB-MUSIC, EOCAB-MUSIC and OCAB-MUSIC when the nested array given in (\ref{nested}) is employed. The number of snapshots and the SNR are considered to be $N=500$ and $0$ dB, respectively. In addition, we consider two sources with equal powers, located at $\theta_1=\ang{20}-\frac{\Delta \theta} {2}$ and $\theta_2=\ang{20}+\frac{\Delta \theta}{2}$. We define the two sources as being resolvable if ${\rm \underset{\Scale[0.5]{i \in \{1,2\}}} {max}}|\hat{\theta}_i-\theta_i|<\frac{\Delta \theta}{2}$ \cite{Kaveh1986}. According to this definition and making use of two-dimensional Chebychev's bound \cite{Lal1955}, the probability of resolution can be lower bounded as
\small
\begin{align}
\label{Eq-ProRes}
&\mathbb{P}({\rm \underset{\Scale[0.5]{i \in \{1,2\}}} {max}}|\hat{\theta}_i-\theta_i|<\frac{\Delta \theta}{2}) \\
&= \mathbb{P}(|\hat{\theta}_1-\theta_1|<\frac{\Delta \theta}{2},  |\hat{\theta}_2-\theta_2|<\frac{\Delta \theta}{2}) \geq \nonumber 1-\frac{2[{\cal E}(\theta_1)+{\cal E}(\theta_2) ]}{\Delta \theta^2}\\
&+\frac{2\sqrt{{\cal E}^2_{\theta_1}+{\cal E}^2_{\theta_2}+2{\cal E}_{\theta_1}{\cal E}_{\theta_2}-4{\cal E}^2_{\theta_1,\theta_2}}}{\Delta \theta ^2}, \nonumber
\end{align}\normalsize
where ${\cal E}(\theta_1)$, ${\cal E}(\theta_2)$ and ${\cal E}(\theta_1,\theta_2)$ are given in \eqref{Eq-mse} and \eqref{Eq-cov}. The analytical expression on the right-hand side of \eqref{Eq-ProRes} enables us to predict the minimum source separation required for achieving a particular probability of resolution. For example, Fig. \ref{Fig:4} shows the predicted values for the minimum source separation to achieve a probability of resolution greater than $0.9$, obtained from \eqref{Eq-ProRes}, for ICAB-MUSIC, OCAB-MUSIC and EOCAB-MUSIC. It is observed that the predicted values of the minimum source separation for ICAB-MUSIC, EOCAB-MUSIC and OCAB-MUSIC, which are respectively $\Delta \theta = \ang{1.2}$, $\Delta \theta = \ang{1.4}$ and $\Delta \theta = \ang{1.5}$, are in a good agreement with the values obtained from the numerical simulations, which are respectively $\Delta \theta = \ang{1.1}$, $\Delta \theta = \ang{1.2}$ and $\Delta \theta = \ang{1.3}$. Additionally, Fig. \ref{Fig:5} demonstrates the resolution performance of EOCAB-MUSIC is superior to that of OCAB-MUSIC while
ICAB-MUSIC outperforms both of them.
\section{Conclusion}
\label{sec:conclusion}
In this paper, we considered the problem of DoA estimation from one-bit measurements received by an SLA. We showed that the idetifiability condition for the DoA estimation problem from one-bit SLA data is equivalent to that for the case when DoAs are estimated from infinite-bit unquantized measurements. Then, we derived a pessimistic approximation of the corresponding CRB. This pessimistic CRB was used as a benchmark for assessing the performance of one-bit DoA estimators. Further, it provides us with valuable insights on the performance limits of DoA estimation from one-bit quantized data. For example, it was shown that the DoA estimation errors in one-bit scenario reduces at the same rate as that of infinite-bit case with respect to the number of samples and, moreover, that the DoA estimation errors in one-bit scenario converges to a constant value by increasing the SNR. We also proposed a new algorithm for estimating DoAs from one-bit quantized data. We investigated the analytical performance of the proposed method through deriving a closed-form expression for the second-order statistics of its asymptotic distribution (for the large number of snapshots) and show that it outperforms the existing algorithms in the literature. Numerical simulations were provided to validate the analytical derivations and corroborate the improvement in estimation performance.
\appendices
\section{Proof of Theorem \ref{Theo-1}}\label{app-A}
We first prove the sufficiency. Assume that $\THETA_0 \in [-\pi/2, \pi/2]^{K \times 1}$ is identifiable from $\YM$. This implies that $f(\YM \mid \THETA_0, \PV, \sigma^2) \neq f(\YM \mid \breve{\THETA}, \breve{\PV}, \breve{\sigma}^2)$ for any arbitrary values of $\breve{\THETA} \neq \THETA_0 \in [-\pi/2, \pi/2]^{K \times 1}$, $\PV \in \mathds{R}_{>0}^{{K \times 1}}$, $\breve{\PV} \in \mathds{R}_{>0}^{{K \times 1}}$, $\sigma^2$ and $\breve{\sigma}^2$.
Hence, considering $\YV(0), \YV(1), \cdots, \YV(N-1)$ are independent and identically distributed with $\YV(t) \sim {\cal CN}(\ZEROV, \RM)$, we have
\small
\begin{align}
\label{Eq-app-A-1}
\AM\!(\THETA_0)\DIAG(\PV)
\AM\!^H\!(\THETA_0)\!+\!\sigma^2\IDM_M \!\neq\! \AM\!(\breve{\THETA})\DIAG(\breve{\PV})
\AM\!^H\!(\breve{\THETA})\!+\!\breve{\sigma}^2\IDM_M,
\end{align}\normalsize
for all $\breve{\THETA} \neq \THETA_0 \in [-\pi/2, \pi/2]^{K \times 1}$, $\PV \in \mathds{R}_{>0}^{{K \times 1}}$, $\breve{\PV} \in \mathds{R}_{>0}^{{K \times 1}}$, $\sigma^2$ and $\breve{\sigma}^2$.

In what follows, we employ the method of proof by contradiction to prove the sufficiency. In particular, we assume that
$\THETA_0 \in [-\pi/2, \pi/2]^{K \times 1}$ is non-identifiable from $\XM$. Hence, there exists a $\breve{\THETA} \neq \THETA_0 \in [-\pi/2, \pi/2]^{K \times 1}$ at which $f(\XM \mid \THETA_0, \widetilde{\PV}, \widetilde{\sigma}^2) = f(\XM \mid \breve{\THETA}, \dot{\PV}, \dot{\sigma}^2)$ for some values of $\widetilde{\PV} \in \mathds{R}_{>0}^{{K \times 1}}$, $\dot{\PV} \in \mathds{R}_{>0}^{{K \times 1}}$, $\widetilde{\sigma}^2$ and $\dot{\sigma}^2$. It is readily clear from assumption {\bf A4} and \eqref{model-eq-6} that $\EX\{\XV(t_1) \XV^H(t_2)\}=\ZEROV$ when $t_1 \neq t_2$. Accordingly, we have
\small
\begin{align}
\label{Eq-app-A-5}
&\EX\left\{\XM\XM^H \!\mid\! \THETA_0, \widetilde{\PV}, \widetilde{\sigma}^2 \right\} = \EX\left\{\XM\XM^H \!\mid\! \breve{\THETA}, \dot{\PV}, \dot{\sigma}^2 \right\},\\
\Rightarrow & \! \sum_{t=0}^{N-1} \EX\{\XV(t)\XV^H(t) \!\mid\! \THETA_0, \widetilde{\PV}, \widetilde{\sigma}^2\} \!=\! \sum_{t=0}^{N-1} \EX\{\XV(t)\XV^H(t) \!\mid\! \breve{\THETA}, \dot{\PV}, \dot{\sigma}^2\}. \nonumber
\end{align}\normalsize
From \eqref{Eq-app-A-5}, \eqref{Eq-arclaw}, \eqref{model-eq-3} and the fact that the arcsine function is one-to-one when its argument is between $-1$ and $1$, it follows that
\small
\begin{align}
\label{Eq-app-A-6}
&\frac{1}{\widetilde{\sigma}^2+\sum_{k=1}^K \tilde{p}_k}\left[\AM(\THETA_0)\DIAG(\widetilde{\PV})
\AM^H(\THETA_0)\!+\!\widetilde{\sigma}^2\IDM_M \right] = \nonumber\\
& \frac{1}{\dot{\sigma}^2+\sum_{k=1}^K \dot{p}_k} \left[\AM(\breve{\THETA})\DIAG(\dot{\PV})
\AM^H(\breve{\THETA})\!+\!\dot{\sigma}^2\IDM_M \right].
\end{align}\normalsize
Considering $\PV = \frac{\widetilde{\PV}}{\widetilde{\sigma}^2+\sum_{k=1}^K \tilde{p}_k}$, $\sigma^2 = \frac{\widetilde{\sigma}^2}{\widetilde{\sigma}^2+\sum_{k=1}^K \tilde{p}_k}$, $\breve{\PV} = \frac{\dot{\PV}}{\dot{\sigma}^2+\sum_{k=1}^K \dot{p}_k}$ and $\breve{\sigma}^2 = \frac{\dot{\sigma}^2}{\dot{\sigma}^2+\sum_{k=1}^K \dot{p}_k}$, we obtain
\small
\begin{align}
\label{Eq-app-A-6-1}
\AM(\THETA_0)\DIAG(\PV)
\AM^H(\THETA_0)+\sigma^2\IDM_M = \AM\!(\breve{\THETA})\DIAG(\breve{\PV})
\AM^H(\breve{\THETA})+\breve{\sigma}^2\IDM_M,
\end{align}\normalsize
which is in contradiction with \eqref{Eq-app-A-1}.
Hence, the initial assumption that $\THETA_0 \in [-\pi/2, \pi/2]^{K \times 1}$ is non-identifiable from $\XM$ cannot be true. This proves the sufficiency.

To show the necessity, let assume that $\THETA_0 \in [-\pi/2, \pi/2]^{K \times 1}$ is non-identifiable from $\YM$. This implies that there exist some $\breve{\THETA} \in [-\pi, \pi]^{q \times 1} \neq \THETA_0$, $\PV$, $\breve{\PV}$, $\sigma^2$ and $\breve{\sigma}^2$ for which $f(\YM \mid \THETA_0, \PV, \sigma^2) = f(\YM \mid \breve{\THETA}, \breve{\PV}, \breve{\sigma}^2)$. Since the true PDF of $\XM$ is obtained from the orthant probabilities of $\YM$, it is readily deduced that $f(\XM \mid \THETA_0, \PV, \sigma^2) = f(\XM \mid \breve{\THETA}, \breve{\PV}, \breve{\sigma}^2)$ as well. This proves that identifiability of $\THETA_0 \in [-\pi/2, \pi/2]^{K \times 1}$ from $\YM$ is a necessary condition for identifiability of $\THETA_0 \in [-\pi/2, \pi/2]^{K \times 1}$ from $\XM$.
\section{Proof of Theorem \ref{Theo-2}}\label{app-B}
We first prove {\bf S1}. Consider arbitrary $\THETA \neq  \in [-\pi/2, \pi/2]^{K \times 1}$ and $\breve{\THETA} \in [-\pi/2, \pi/2]^{K \times 1}$ such that $\THETA_k \neq \breve{\THETA}$. Moreover, let $\AM_v(\THETA)$ be the steering matrix of a contiguous ULA with $v$ elements located at $(0 ,\frac{\lambda}{2}, \cdots, (v-1)\frac{\lambda}{2})$. Considering the fact that $\AM_v(\THETA)$ is a Vandermonde matrix, if $K \leq v-1$, it follows from Caratheodory-Fejer-Pisarenko decomposition \cite{caratheodory1911zusammenhang} that
\small
\begin{align}
\label{Eq-App-B-1}
\AM_v(\THETA) \DIAG(\PV) \AM_v^H\!(\THETA) \!+\! \sigma^2 \IDM_v \!\neq\! \AM_v\!(\breve{\THETA}) \DIAG(\breve{\PV})\AM_v^H\!(\breve{\THETA}) \!+\! \breve{\sigma}^2 \IDM_v,
\end{align}\normalsize
for any arbitrary values of $\PV \in \mathds{R}_{>0}^{{K \times 1}}$, $\breve{\PV} \in \mathds{R}_{>0}^{{K \times 1}}$, $\sigma^2$ and $\breve{\sigma}^2$. From \cite[Eq. (113)]{SedighiTSP2019}, vectorizing both sides of \eqref{Eq-App-B-1} leads to
\small
\begin{align}
\label{Eq-app-B-2}
\TM{'} \AM_{\vartheta}(\THETA)\PV
+\sigma^2\TM{'}\EV{'} \neq
\TM{'}\AM_{\vartheta}(\breve{\THETA})\breve{\PV}
+\breve{\sigma}^2\TM{'}\EV{'},
\end{align}\normalsize
where $\mathbf{A}_{\vartheta}(\THETA) \in \mathds{C}^{(2v-1) \times K}$ denotes the steering matrix corresponding to the contiguous ULA segment of the difference co-array, $\mathbf{T}' \in \{0,1\}^{v^2 \times (2v-1)}$ is a selection matrix defined in \cite[Eq. (114)]{SedighiTSP2019} and $\EV{'} \in \{0,1\}^{(2v-1) \times 1}$ is a column vector with $[\EV{'}]_i=\delta[i-v]$. Considering $\TM'$ is full-column rank \cite{SedighiTSP2019}, multiplying both sides of \eqref{Eq-app-B-2} by $\TM'^{\dagger}$ and then moving all the terms to one side of the equation yields
\small
\begin{align}
\label{Eq-app-A-7}
\AM_{\vartheta}(\THETA)\PV -\AM_{\vartheta}(\breve{\THETA})\breve{\PV}
+(\sigma^2
-\breve{\sigma}^2)\EV{'} \neq \ZEROV.
\end{align}\normalsize
It follows from $\breve{\THETA} \!\neq\! \THETA_0$ that $\breve{\THETA}$ could differs from $\THETA_0$ at $q$ DoAs for some integer $q \!\in\! [1,K]$. Noting this fact, \eqref{Eq-app-A-7} is simplified to
\small
\begin{align}
\label{Eq-app-B-3}
\begin{bmatrix} \AM_{\vartheta}(\THETA) & \AM_{\vartheta}(\ddot{\THETA}) & \EV{'} \end{bmatrix} \begin{bmatrix}
\PV - \breve{\PV} \odot \VAREPSILON \\ -\ddot{\PV} \\  \sigma^2 - \breve{\sigma}^2
\end{bmatrix} \neq \ZEROV,
\end{align}\normalsize
where $\ddot{\THETA} \in [-\pi, \pi]^{q \times 1}$ consists of those elements of $\breve{\THETA}$ which do not intersect with those in $\THETA$,  $\ddot{\PV} \in \mathds{R}_{>0}^{q \times 1}$ contains those elements of $\breve{\PV}$ corresponding to $\ddot{\THETA}$ and
\small
\begin{align}
\label{Eq-app-A-9}
 [\VAREPSILON]_i = \left\{\begin{array}{cc}
           1, & [\THETA]_i = [\breve{\THETA}]_i, \\
            0, & \text{otherwise}.
          \end{array}
\right.
\end{align}\normalsize
Considering that \small$\begin{bmatrix}\AM_{\vartheta}(\THETA) \!&\! \AM_{\vartheta}(\breve{\THETA}) \!&\! \EV^{'} \end{bmatrix} \in \mathds{C}^{(2v-1) \times (2K+1)}$\normalsize~ is a sub-matrix of \small$\begin{bmatrix} \AM_d(\THETA) \!&\! \AM_d(\ddot{\THETA}) \!&\! \EV \end{bmatrix} \in \mathds{C}^{(2D-1) \times (2K+1)}$\normalsize, obtained from $2v-1$ rows of \small$\begin{bmatrix}\AM_d(\THETA) \!&\! \AM_d(\ddot{\THETA}) \!&\! \EV\end{bmatrix}$\normalsize, it follows from \eqref{Eq-app-B-3} that
\small
\begin{align}
\label{Eq-app-B-3-1}
&\begin{bmatrix} \AM_d(\THETA) & \AM_d(\ddot{\THETA}) & \EV \end{bmatrix} \begin{bmatrix}
\PV - \breve{\PV} \odot \VAREPSILON \\ -\ddot{\PV} \\  \sigma^2 - \breve{\sigma}^2
\end{bmatrix} \neq \ZEROV,
\\
\label{Eq-app-B-3-2}
\Rightarrow & \AM_d(\THETA)\PV -\AM_d(\breve{\THETA})\breve{\PV}
+(\sigma^2
-\breve{\sigma}^2)\EV \neq \ZEROV.
\end{align}\normalsize
Multiplying \eqref{Eq-app-B-3-2} by $\JM$ and exploiting \eqref{model-eq-3} and \eqref{model-eq-4}, after some algebraic manipulations, we obtain
\small
\begin{align}
 &\VE(\AM(\THETA) \DIAG(\PV) \AM^H(\THETA) + \sigma^2 \IDM_M) \nonumber\\
 &\neq \VE(\AM(\breve{\THETA}) \DIAG(\breve{\PV})\AM^H(\breve{\THETA})     + \breve{\sigma}^2 \IDM_M),
\end{align}\normalsize
which in turn implies that
\small
\begin{align}
\label{Eq-App-B-4}
\AM\!(\THETA) \DIAG(\PV) \AM^H\!(\THETA) \!+\! \sigma^2 \IDM_M \!\neq\! \AM\!(\breve{\THETA}) \DIAG(\breve{\PV})\AM^H\!(\breve{\THETA}) \!+\! \breve{\sigma}^2 \IDM_M,
\end{align}\normalsize
for all $\THETA \neq \breve{\THETA}  \in [-\pi/2, \pi/2]^{K \times 1}$, $\PV \in \mathds{R}_{>0}^{{K \times 1}}$, $\breve{\PV} \in \mathds{R}_{>0}^{{K \times 1}}$, $\sigma^2$ and $\breve{\sigma}^2$. Considering $\YV(0), \YV(1), \cdots, \YV(N-1)$ are independent and identically distributed with $\YV(t) \sim {\cal CN}(\ZEROV, \RM)$, it follows from \eqref{Eq-App-B-4} that $f(\YM \mid \THETA_0, \PV, \sigma^2) \neq f(\YM \mid \breve{\THETA}, \breve{\PV}, \breve{\sigma}^2)$ for any arbitrary values of $\THETA \neq \breve{\THETA} \in [-\pi/2, \pi/2]^{K \times 1}$, $\PV \in \mathds{R}_{>0}^{{K \times 1}}$, $\breve{\PV} \in \mathds{R}_{>0}^{{K \times 1}}$, $\sigma^2$ and $\breve{\sigma}^2$ if $K \leq v-1$. Now, from Theorem \ref{Theo-1}, we conclude that  $f(\XM \mid \THETA_0, \PV, \sigma^2) \neq f(\XM \mid \breve{\THETA}, \breve{\PV}, \breve{\sigma}^2)$. This completes the proof of {\bf S1}.

We now prove {\bf S2}. We know from Lemma \ref{Lem-1} that the FIM is singular for any value of $\THETA \in [-\pi/2, \pi/2]^{K \times 1}$ if $K \geq D$. This means that the problem is not even locally indentifiable at any $\THETA$ \cite{rothenberg1971identification}. Since the local identifiablity is a necessary condition for the identifiablity any particular point, the problem is not identifiable for any $\THETA$.
\section{Proof of Lemma \ref{Lem-1}}\label{app-C}
Let $\RM_{\XV}^r$ and $\overline{\RM}^r$ denote the equivalent real representation for $\RM_{\XV}$ and $\overline{\RM}$, respectively, given as
\small
\begin{align}
   \RM_{\XV}^r = \begin{bmatrix}
       \Re\{\RM_{\XV}\} & -\Im\{\RM_{\XV}\} \\
       \Im\{\RM_{\XV}\} & \Re\{\RM_{\XV}\}
    \end{bmatrix},~~
  \overline{\RM}^r =  \begin{bmatrix}
       \Re\{\overline{\RM}\} & -\Im\{\overline{\RM}\} \\
       \Im\{\overline{\RM}\} & \Re\{\overline{\RM}\}.
    \end{bmatrix}
\end{align}\normalsize
Making use of \eqref{Eq-arclaw} and Taylor expansion of arcsine function, we have
\small
\begin{align}
\label{Eq-app-C-0}
\RM_{\XV}^r = & \frac{2}{\pi} \arcsin(\overline{\RM}^r) = \overline{\RM}^r + \frac{1}{6} \overline{\RM}^r \odot \overline{\RM}^r \odot \overline{\RM}^r \nonumber \\
&+\frac{3}{40} \overline{\RM}^r \odot \overline{\RM}^r \odot \overline{\RM}^r \odot \overline{\RM}^r \odot \overline{\RM}^r + \cdots \nonumber\\
=& \sum_{n=0}^{\infty} \frac{(2n)!}{(2^n n!)^2 (2n+1)} \underbrace{\overline{\RM}^r \odot \overline{\RM}^r \odot \cdots \odot \overline{\RM}^r}_{2n+1~{\rm times}}.
\end{align}\normalsize
It is clear from \eqref{Eq-CRB-1} that $\overline{\RM}$ is positive definite, and so is $\overline{\RM}^r$. Further, it follows from the Schur product theorem \cite[Theorem 3.1]{styan1973hadamard}, which establishes that the Hadamard product of two positive-definite matrices is also a positive-definite matrix, that the $2n+1$ times Hadamard products of $\overline{\RM}^r$ by itself is also positive definite for any integer $n \in [0, \infty)$. Hence, it follows from \eqref{Eq-app-C-0} that $\RM_{\XV}^r$ is obtained from a weighted sum of positive definite matrices, and thus it is positive definite. Evidently, $\RM_{\XV}$ is also positive definite. This in turn indicates non-singularity of $(\RM_{\XV}^{-T} \otimes \RM_{\XV}^{-1})$. Hence, since $\JM$ is also full-column rank \cite{Liu2017}, we easily conclude that $\JM^H (\RM_{\XV}^{-T} \otimes \RM_{\XV}^{-1}) \JM$ is full rank. This implies that ${\cal I}_w(\VARRHO)$ is non-singular if and only if \small$\begin{bmatrix}
\GM & \VM \end{bmatrix} \in \mathds{R}^{(2D-1) \times 2K}$\normalsize is full-column rank. In other words, ${\cal I}_w(\VARRHO)$ is non-singular if and only if
\small
\begin{align}
\label{Eq-app-C-1}
\begin{bmatrix}
\GM & \VM \end{bmatrix} \begin{bmatrix}
\CV_1 \\ \CV_2
\end{bmatrix}  \neq \ZEROV,
\end{align}\normalsize
for any arbitrary non-zero $\CV = [\CV_1^T, \CV_2^T]^T \in \mathds{C}^{2K \times 1}$.
Inserting \eqref{Eq-CRB-9} and \eqref{Eq-CRB-10} into \eqref{Eq-app-C-1} leads to
\small
%
%
\begin{align}
 \begin{bmatrix} \DELTAM & \DIGAMMAM \end{bmatrix}  \begin{bmatrix}
\widetilde{\CV}_1 \\ \CV_2
\end{bmatrix} \neq \ZEROV,
\end{align}\normalsize
where $\widetilde{\CV}_1\!=\!\J \pi \boldsymbol{\Phi}(\THETA) \DIAG(\overline{\PV}) \CV_1$. This completes the proof.
\section{Proof of Theorem \ref{Theo-3}}\label{app-D}
We know from Appendix \ref{app-C} that $\MM = \JM^H (\RM_{\XV}^{-T} \otimes \RM_{\XV}^{-1}) \JM$ is positive-definite. Hence, \eqref{Eq-CRB-8} can be rewritten as
\small
\begin{align}
    \label{Eq-app-D-1}
 {\cal I}_w(\VARRHO) &= N \begin{bmatrix} \GM^H \MM^{\frac{1}{2}} \\ \VM^H \MM^{\frac{1}{2}}
 \end{bmatrix} \begin{bmatrix}  \MM^{\frac{1}{2}}\GM & \MM^{\frac{1}{2}}\VM
 \end{bmatrix}  \nonumber\\
 &= N \begin{bmatrix}
\GM^H \MM \GM & \GM^H \MM \VM \\ \VM^H \MM \GM & \VM^H \MM \VM
 \end{bmatrix}.
\end{align}\normalsize
The $CRB_w(\THETA)$ is then obtained by block-wise inversion as follows:
\small
\begin{align}
\label{Eq-app-D-3}
CRB_w(\THETA)&\!=\!\frac{1}{N}\left(\GM^H \MM \GM \!-\! \GM^H \MM \VM  \left(\VM^H \MM \VM \right)^{-1} \VM^H \MM  \GM \right)^{-1} \nonumber\\
& \!=\! \frac{1}{N}\left(\GM^H \MM^{\frac{1}{2}} \Pi^{\perp}_{\MM^{\frac{1}{2}} \VM} \MM^{\frac{1}{2}}  \GM \right)^{-1} .
\end{align}\normalsize
The facts that $\GM = \J \pi \DIAG(\DV) \OMEGAM \PHIM(\THETA) \DIAG(\overline{\PV})$ and $\RM_{\XV} = \frac{2}{\pi} \asin(\overline{\RM})$ will lead to \eqref{Eq-Pes-CRB}. In addition, It follows from ${\cal I}(\varrho) \succeq {\cal I}_w(\VARRHO)$ that $CRB(\THETA) \preceq CRB_w(\THETA)$.
\section{Proof of Theorem \ref{Theo-5}}\label{app-E}
Recalling $\overline{p}_k = \frac{p_k}{\sigma^2+\sum_{k=1}^K p_k}$ and assuming that all sources have equal power $p$, we have
\small
\begin{align}
\label{Eq-app-E-1}
\lim_{SNR \to \infty} \overline{p}_k =  \lim_{SNR \to \infty} \frac{SNR}{K \times SNR +1 } = \frac{1}{K}.
\end{align}\normalsize
Making use of \eqref{Eq-app-E-1}, it can be readily shown that
\small
\begin{align}
\label{Eq-app-E-2}
\lim_{SNR \to \infty} \overline{\RM} &= \frac{1}{K} \AM(\THETA) \AM^H(\THETA) + (1-\frac{1}{K}) \IDM_M.
\end{align}\normalsize
The above equation implies that $\lim_{SNR \to \infty} \overline{\RM}$ is a positive-definite matrix independent of the SNR. Further, it follows from \eqref{Eq-app-E-1} that
\small
\begin{align}
\label{Eq-app-E-3}
&\lim_{SNR \to \infty}\DIAG(\overline{\PV}) = \frac{1}{K} \IDM_K,\\
\label{Eq-app-E-4}
&\lim_{SNR \to \infty} \HV =  \bigg[\frac{1}{\sqrt{1-\frac{|\Re\{\sum_{k=1}^K e^{-\J \pi \sin \theta_k \ell_{D-1}}\}|^2}{K^2}}}, \cdots, 0, \nonumber\\
&~~~~~ \cdots, \frac{1}{\sqrt{1-\frac{|\Re\{\sum_{k=1}^K e^{\J \pi \sin \theta_k \ell_{D-1}}\}|^2}{K^2}}}\bigg]^T,\\
\label{Eq-app-E-5}
&\lim_{SNR \to \infty} \overline{\HV} =  \bigg[\frac{1}{\sqrt{1-\frac{|\Im\{\sum_{k=1}^K e^{-\J \pi \sin \theta_k \ell_{D-1}}\}|^2}{K^2}}}, \cdots, 0, \nonumber\\
&~~~~~ \cdots, \frac{1}{\sqrt{1-\frac{|\Im\{\sum_{k=1}^K e^{\J \pi \sin \theta_k \ell_{D-1}}\}|^2}{K^2}}}\bigg]^T.
\end{align}\normalsize
Substituting \eqref{Eq-app-E-3}, \eqref{Eq-app-E-4} and \eqref{Eq-app-E-5} back into \eqref{Eq-CRB-9} and \eqref{Eq-CRB-10} indicates that $\lim_{SNR \to \infty} \begin{bmatrix}
 \GM & \VM
\end{bmatrix}$ is a full-column rank matrix independent of the SNR. Hence, recalling \eqref{Eq-CRB-8}, we can conclude that $\lim_{SNR \to \infty}{\cal I}_w(\varrho)$ is positive-definite and independent of the SNR. This in turn implies that $\lim_{SNR \to \infty}CRB_w(\THETA)$, which is Schur complement of $\lim_{SNR \to \infty}{\cal I}_w(\varrho)$, is also positive-definite and independent of the SNR. This completes the proof.
\section{Proof of Lemma \ref{lem-new-2}}
\label{app-new-1}
We start with showing that $\PSIM$ is full rank. Making use of relations \small$\det(\begin{bmatrix} \CM_1 & \CM_2\\ \CM_3 & \CM_4\end{bmatrix}) = \det(\CM_1) \det(\CM_4 - \CM_3 \CM_1^{-1} \CM_2) $\normalsize, we obtain
\small
\begin{align}
\det(\PSIM) = \det(\IDM_{D-1}) \det(2\J \IDM_{D-1}) = (2\J)^{D-1} \neq 0,
\end{align}\normalsize
which implies full rankness of $\PSIM$.

Next, we proceed with proving that $\overline{\JM}$ is full rank. Let $\ddot{\JM}$ denote the matrix obtained after removing the $D$-th column from $\JM$. $\ddot{\JM}$ is full column rank since its columns are a sub-set of the columns of the full-column-rank matrix $\JM$ \cite{Liu2017}. Further, for $1 \leq i \leq M$, it is readily confirmed that the $((i-1)M+1)$-th row of $\VE(\LM_n)$ as well as $\VE(\LM_n^T)$ equals the $i$-th diagonal element of $\LM_n$, which is obviously zero for $n \neq 0$ according to the definition given after \eqref{model-eq-5}. Given \eqref{model-eq-5}, this in turn implies that the rows of $\ddot{\JM}$ with indices $(i-1)M+1$, for all $1 \leq i \leq M$,
are zero vectors. As a result, the matrix obtained by removing these rows from $\ddot{\JM}$, i.e., $\overline{\JM}$, has the same column rank as $\ddot{\JM}$. This completes the proof.

Finally, we show that $\FM$ is full rank. It follows from the fact that $\overline{\EV}_p^T \overline{\EV}_q = 0$ for $p \neq q$ that
\small
\begin{align}
   &(\overline{\EV}_i^T \otimes \overline{\EV}_j^T \pm \overline{\EV}_j^T \otimes \overline{\EV}_i^T) (\overline{\EV}_p \otimes \overline{\EV}_q \pm \overline{\EV}_q \otimes \overline{\EV}_p) =  \overline{\EV}_i^T \overline{\EV}_p \otimes \overline{\EV}_j^T \overline{\EV}_q \nonumber\\
   &\pm \overline{\EV}_i^T \overline{\EV}_q \otimes \overline{\EV}_j^T \overline{\EV}_p \pm \overline{\EV}_j^T \overline{\EV}_p \otimes \overline{\EV}_i^T \overline{\EV}_q \pm \overline{\EV}_j^T \overline{\EV}_q \otimes \overline{\EV}_i^T \overline{\EV}_p = 0.
\end{align}\normalsize
for $1 \leq i < j \leq M$ and $1 \leq p < q \leq M$ when either $p$ or $q$ differs from $i$ and $j$. In addition, in case $i=p$ and $j=q$, we have
\small
\begin{align}
&(\overline{\EV}_i^T \otimes \overline{\EV}_j^T + \overline{\EV}_j^T \otimes \overline{\EV}_i^T) (\overline{\EV}_i \otimes \overline{\EV}_j - \overline{\EV}_j \otimes \overline{\EV}_i) =  \overline{\EV}_i^T \overline{\EV}_i \otimes \overline{\EV}_j^T \overline{\EV}_j \nonumber\\
   &- \overline{\EV}_i^T \overline{\EV}_j \otimes \overline{\EV}_j^T \overline{\EV}_i + \overline{\EV}_j^T \overline{\EV}_i \otimes \overline{\EV}_i^T \overline{\EV}_j - \overline{\EV}_j^T \overline{\EV}_j \otimes \overline{\EV}_i^T \overline{\EV}_i = 0.
\end{align}\normalsize
It is also observed that, for $1 \leq i \leq M$ and $1 \leq p < q \leq M$, the $((i-1)M+1)$-th element of $\overline{\EV}_p^T \otimes \overline{\EV}_q^T \pm \overline{\EV}_p^T \otimes \overline{\EV}_q^T$ is equal to the $i$-th diagonal element of $\overline{\EV}_p \overline{\EV}_q^T \pm \overline{\EV}_p\overline{\EV}_q^T$, which is obviously zero for $p \neq q$. Consequently, the row vectors obtained by removing the elements with indices $(i-1)M+1$ for all $1 \leq i \leq M$ from $\overline{\EV}_p^T \otimes \overline{\EV}_q^T + \overline{\EV}_q^T \otimes \overline{\EV}_p^T$ and $\overline{\EV}_p^T \otimes \overline{\EV}_q^T - \overline{\EV}_q^T \otimes \overline{\EV}_p^T$ will be still orthogonal with each other. Hence, it is deduced that the square matrix $\FM$ has orthogonal rows, thereby being full rank.
\section{Proof of Lemma \ref{lem-2}}
\label{App-F}
Let define $E(\theta) = \AV_v^H(\theta) \widehat{\UM}_n \widehat{\UM}^H_n \AV_v(\theta)$ and $\breve{E}(\theta) = \AV_v^H(\THETA) \UM_n \UM^H_n \AV_v(\THETA)$ where $\widehat{\UM}_n$ and $\UM_n$ consist of, respectively, the eigenvectors of $\widehat{\overline{\RM}}_v$ and $\AM_v(\THETA) \DIAG(\overline{\PV}) \AM_v^H(\THETA) + \overline{\sigma}^2 \IDM_v$ corresponding to their $v-K$ smallest eigenvalues with $K \leq v-1$. We know that the elements of $\widehat{\THETA}$ are equal to the minimizers of $E(\theta)$. Defining $E_n = {\underset{\theta} {\rm sup}}|E(\theta) - \breve{E}(\theta)|$, we have
\small
\begin{align}
\label{Eq-app-F-1}
 E_n &= {\underset{\theta} {\rm sup}} \left| \AV_v^H(\THETA) (\widehat{\UM}_n \widehat{\UM}^H_n-\UM_n \UM^H_n) \AV_v(\THETA) \right|
 \nonumber\\
 &= {\underset{\theta} {\rm sup}} \left|\left(\AV_v^T(\THETA) \otimes \AV_v^H(\THETA) \right) \VE(\widehat{\UM}_n \widehat{\UM}^H_n-\UM_n \UM^H_n) \right| \nonumber\\
&\leq \|\AV_v^T(\THETA) \otimes \AV_v^H(\THETA)\|_2 \|\VE(\widehat{\UM}_n \widehat{\UM}^H_n-\UM_n \UM^H_n)\|_2 \nonumber\\
&= v^2 \|\VE(\widehat{\UM}_n \widehat{\UM}^H_n-\UM_n \UM^H_n)\|_2.
\end{align}\normalsize
It follows from \eqref{Eq-Est-21} that $\lim_{N \to \infty} \widehat{\UM}_n \widehat{\UM}^H_n = \UM_n \UM^H_n$. Hence, $E_n \to 0$ as $N \to \infty$. This implies that $E(\theta)$ converges uniformly to $\breve{E}(\theta)$ as $N \to \infty$.
Thus, the minimizers of $E(\theta)$, i.e., the elements of $\widehat{\THETA}$, converge to the minimizers of $\breve{E}(\theta)$, i.e., $\theta_1, \theta_2, \cdots, \theta_K$, as $N \to \infty$. This completes the proof.
\section{Proof of Theorem \ref{Theo-6}}
\label{app-G}
Considering the consistency of $\widehat{\THETA}$ and following the same arguments as in \cite[App. B]{Wang2017}, for sufficiently large $N$, the asymptotic estimation error expression for EOCAB-MUSIC is given by
\begin{align}
\label{Eq-app-G-1}
 \hat{\theta}_k - \theta_k =  - \frac{ \Re\{\ZV^T_k \TM \JM^{\dagger} \Delta \overline{\RV}\} }{\pi \overline{p}_k q_k \cos(\theta_k)}   ,
\end{align}
where $\Delta \overline{\RV} = \widehat{\overline{\RV}} - \overline{\RV}$ and $\mathbf{T}=
\begin{bmatrix}
   \mathbf{T}_v^T & \mathbf{T}_{v-1}^T & \cdots &
   \mathbf{T}_1^T
\end{bmatrix}^T \in \mathds{C}^{v^2 \times (2D-1)}$.
From \eqref{Eq-app-G-1}, the covariance of the asymptotic distribution (as $N \to \infty$) of the DoA estimation errors is given by
\small
\begin{align}
\label{Eq-app-G-2}
{\cal E}_{\theta_{k_1}, \theta_{k_2}} &= \EX\{ (\hat{\theta}_{k_1} - \theta_{k_1})(\hat{\theta}_{k_2} - \theta_{k_2}) \} \nonumber\\
&= \frac{ \EX\left\{ \Re\{\ZV^T_{k_1}  \TM \JM^{\dagger} \Delta \overline{\RV} \} \Re\{\ZV^T_{k_2}  \TM \JM^{\dagger} \Delta \overline{\RV} \} \right\} }{\pi^2 \overline{p}_{k_1} \overline{p}_{k_2} q_{k_1} q_{k_2} \cos(\theta_{k_1}) \cos(\theta_{k_2})}.
\end{align}\normalsize
Making use of the identity $\Re\{\CV_1^H \CV_2\} \Re\{\CV_3^H \CV_2\} = \frac{1}{2}\Re\{\CV_1^H \CV_2 \CV_2^H \CV_3 +  \CV_1^H \CV_2 \CV_2^T \CV_3^*\}$, we obtain
\small
\begin{align}
\label{Eq-app-G-3}
&\EX\left\{ \Re\{\ZV^T_{k_1}  \TM \JM^{\dagger} \Delta \overline{\RV} \} \Re\{\ZV^T_{k_2}  \TM \JM^{\dagger} \Delta \overline{\RV} \} \right\}  =  \nonumber\\
&\frac{1}{2}\EX\big\{ \Re\{ \ZV^T_{k_1}  \TM \JM^{\dagger} \Delta \overline{\RV} \Delta \overline{\RV}^H \JM^{\dagger H} \TM^H \ZV_{k_2}^*\} \nonumber\\
&+ \Re\{ \ZV^T_{k_1}  \TM \JM^{\dagger} \Delta \overline{\RV} \Delta \overline{\RV}^T \JM^{\dagger H} \TM^H \ZV_{k_2}  \} \big\}. 
\end{align}\normalsize
The matrix $\MAT_{M,M}(\JM^{\dagger H} \TM^H \ZV_k)$ is Hermitian \cite[Lemma 6]{Wang2017}, thereby
\small
\begin{align}
\label{Eq-app-G-4}
 \JM^{\dagger H} \TM^H \ZV_k^* = \KM_M \JM^{\dagger H} \TM^H \ZV_k.
\end{align}\normalsize
where $\KM_M \in \{0,1\}^{M^2 \times M^2}$ is the commutation matrix defined as $\VE(\CM^T) = \KM_M \VE(\CM)$ for any arbitrary matrix $\CM$ \cite{Magnus2007}. In addition, since $\overline{\RM}^H = \overline{\RM}$, we have
\small
\begin{align}
\label{Eq-app-G-5}
 \Delta \overline{\RV}^T =   \Delta \overline{\RV}^H \KM_M^H.
\end{align}\normalsize
Inserting \eqref{Eq-app-G-4} and \eqref{Eq-app-G-5} into \eqref{Eq-app-G-3} and using the fact that $\KM_M = \KM_M^H = \KM_M^{-1}$, we obtain
\small
\begin{align}
\label{Eq-app-G-6}
&\EX\left\{ \Re\{\ZV^T_{k_1}  \TM \JM^{\dagger} \Delta \overline{\RV} \} \Re\{\ZV^T_{k_2}  \TM \JM^{\dagger} \Delta \overline{\RV} \} \right\} \nonumber\\
&= \EX\{\Re\{ \ZV^T_{k_1}  \TM \JM^{\dagger} \Delta \overline{\RV} \Delta \overline{\RV}^H \JM^{\dagger H} \TM^H \ZV_{k_2}^* \}\}.
\end{align}\normalsize
Recalling \eqref{Eq-Est-18} and \eqref{Eq-Est-4}, we have
\small
\begin{align}
\label{Eq-app-G-7}
 \TM \JM^{\dagger} \Delta \overline{\RV}
&=
\TM \begin{bmatrix} \ZEROV & \IDM_{D-1} & -\J \IDM_{D-1} \\
0 & \ZEROV & \ZEROV \\
\ZEROV & \IDM_{D-1} & \J \IDM_{D-1} \end{bmatrix} \begin{bmatrix} 0 \\ \widehat{\PHI} - \PHI \end{bmatrix}.
\end{align}\normalsize
Additionally, from \cite[Eq. (114) and Eq. (116)]{SedighiTSP2019}, we know
\footnotesize
\begin{align}\label{Eq-app-G-8}
&\TM= \\
&\big[ \ZEROV_{v^2 \times (D-v)}, \VE (\overline{\LM}^T_{v-1}), \cdots, \VE (\overline{\LM}_{0}), \cdots, \VE (\overline{\LM}_{v-1}), \ZEROV_{v^2 \times (D-v)} \big],\nonumber
\end{align}\normalsize
where
{\footnotesize{
$
[\overline{\LM}_n]_{p,q}=\left\{\begin{array}{cc}
           1, & \text{if}~~ p-q=n, \\
            0, & \text{otherwise}.
          \end{array}
\right..
$}}
Substituting \eqref{Eq-app-G-8} into \eqref{Eq-app-G-7} yields
\small
\begin{align}
\label{Eq-app-G-9}
   \TM \JM^{\dagger} \Delta \overline{\RV} = \overline{\TM} \PSIM (\widehat{\PHI} - \PHI),
\end{align}\normalsize
where
\begin{align}
\label{Eq-app-G-9-1}
&\Scale[0.8]{\hspace{-1mm}\overline{\TM}}= \\
&\hspace{-1mm}\Scale[0.79]{\big[ \ZEROV_{v^2 \times (D-v)}, \VE (\overline{\LM}^T_{v-1}), \cdots, \VE (\overline{\LM}^T_{-1}), \VE (\overline{\LM}_{1}), \cdots, \VE (\overline{\LM}_{v-1}), \ZEROV_{v^2 \times (D-v)} \big].} \nonumber
\end{align}
Inserting \eqref{Eq-app-G-9} into \eqref{Eq-app-G-6} gives
\small
\begin{align}
\label{Eq-app-G-10}
&\EX\left\{ \Re\{\ZV^T_{k_1}  \TM \JM^{\dagger} \Delta \overline{\RV} \} \Re\{\ZV^T_{k_2}  \TM \JM^{\dagger} \Delta \overline{\RV} \} \right\} \nonumber\\
&= \Re\{ \ZV^T_{k_1}  \overline{\TM} \EX\{\PSIM(\widehat{\PHI} - \PHI) (\widehat{\PHI} - \PHI)^H\PSIM^H\} \overline{\TM}^H \ZV_{k_2}^* \}\}.
\end{align}\normalsize
As a result, for sufficiently large $N$, using a first-order perturbation expansion leads to
\small
\begin{align}
\label{Eq-app-G-11}
 &\EX\{\PSIM(\widehat{\PHI} - \PHI) (\widehat{\PHI} - \PHI)^H\PSIM^H\} \simeq \\
 &\left( \overline{\JM}^H \FM^H \DIAG(\BV) \FM^{-H} \SIGMAM^{-1} \FM^{-1} \DIAG(\BV) \FM \overline{\JM} \right)^{-1} \nonumber\\
&\times \overline{\JM}^H \FM^H \DIAG(\BV) \FM^{-H} \SIGMAM^{-1} \FM^{-1} \DIAG(\BV) \FM \EX\{\widetilde{\ddot{\RV}}\widetilde{\ddot{\RV}}^H \} \nonumber\\
&\times \FM^H \DIAG(\BV) \FM^{-H} \SIGMAM^{-1} \FM^{-1} \DIAG(\BV) \FM \overline{\JM} \nonumber\\
&\times \left( \overline{\JM}^H \FM^H \DIAG(\BV) \FM^{-H} \SIGMAM^{-1} \FM^{-1} \DIAG(\BV) \FM \overline{\JM} \right)^{-1} - \PSIM \PHI \PHI^H \PSIM^H, \nonumber
\end{align}\normalsize
where $\SIGMAM = \SIGMAM(\GAMMA)$ given in Appendix K (kindly refer to the supplementary document) and $[\BV]_n =\frac{1}{\sqrt{1-|[\GAMMA]_n|^2}}$ for $1 \leq n \leq M^2-M$. It remains to compute $\EX\{\widetilde{\ddot{\RV}}\widetilde{\ddot{\RV}}^H \}$. Making use of the relation $\widetilde{\ddot{\RV}} = \sine(\frac{\pi}{2} \widehat{\ddot{\RV}}_{\XV})$, we obtain
\small
\begin{align}
\label{Eq-app-G-12}
&\EX\{[\widetilde{\ddot{\RV}}]_p[\widetilde{\ddot{\RV}}]_q^*\} = \nonumber\\
&\frac{1}{4} \EX\bigg\{ e^{\frac{\J\pi \left(\Re\{[\widehat{\ddot{\RV}}_{\XV}]_p\}-\Re\{[\widehat{\ddot{\RV}}_{\XV}]_q\}\right)}{2} } + e^{-\frac{\J\pi \left(\Re\{[\widehat{\ddot{\RV}}_{\XV}]_p\}-\Re\{[\widehat{\ddot{\RV}}_{\XV}]_q\}\right)}{2} } \nonumber\\
&- e^{\frac{\J\pi \left(\Re\{[\widehat{\ddot{\RV}}_{\XV}]_p\}+\Re\{[\widehat{\ddot{\RV}}_{\XV}]_q\}\right)}{2} } - e^{-\frac{\J\pi \left(\Re\{[\widehat{\ddot{\RV}}_{\XV}]_p\}-\Re\{[\widehat{\ddot{\RV}}_{\XV}]_q\}\right)}{2} } \nonumber\\
& + e^{\frac{\J\pi \left(\Im\{[\widehat{\ddot{\RV}}_{\XV}]_p\}-\Im\{[\widehat{\ddot{\RV}}_{\XV}]_q\}\right)}{2} } + e^{-\frac{\J\pi \left(\Im\{[\widehat{\ddot{\RV}}_{\XV}]_p\}-\Im\{[\widehat{\ddot{\RV}}_{\XV}]_q\}\right)}{2} } \nonumber\\
&- e^{\frac{\J \pi \left(\Im\{[\widehat{\ddot{\RV}}_{\XV}]_p\}+\Im\{[\widehat{\ddot{\RV}}_{\XV}]_q\}\right)}{2} } - e^{- \frac{ \J \pi \left(\Im\{[\widehat{\ddot{\RV}}_{\XV}]_p\}+\Im\{[\widehat{\ddot{\RV}}_{\XV}]_q\}\right)}{2} }\bigg\} \nonumber\\
& + \frac{\J}{4} \EX\bigg\{ e^{\frac{\J \pi \left(\Im\{[\widehat{\ddot{\RV}}_{\XV}]_p\}-\Re\{[\widehat{\ddot{\RV}}_{\XV}]_q\}\right)}{2} } + e^{\frac{-\J \pi \left(\Im\{[\widehat{\ddot{\RV}}_{\XV}]_p\}-\Re\{[\widehat{\ddot{\RV}}_{\XV}]_q\}\right)}{2} } \nonumber\\
&- e^{\frac{\J \pi \left(\Im\{[\widehat{\ddot{\RV}}_{\XV}]_p\}+\Re\{[\widehat{\ddot{\RV}}_{\XV}]_q\}\right)}{2} } - e^{\frac{-\J \pi \left(\Im\{[\widehat{\ddot{\RV}}_{\XV}]_p\}+\Re\{[\widehat{\ddot{\RV}}_{\XV}]_q\}\right)}{2} } \nonumber\\
& - e^{\frac{\J \pi \left(\Re\{[\widehat{\ddot{\RV}}_{\XV}]_p\}-\Im\{[\widehat{\ddot{\RV}}_{\XV}]_q\}\right)}{2} } - e^{- \frac{\J \pi \left(\Re\{[\widehat{\ddot{\RV}}_{\XV}]_p\}-\Im\{[\widehat{\ddot{\RV}}_{\XV}]_q\}\right)}{2} } \nonumber\\
&+ e^{\frac{\J \pi \left(\Re\{[\widehat{\ddot{\RV}}_{\XV}]_p\}+\Im\{[\widehat{\ddot{\RV}}_{\XV}]_q\}\right)}{2}} + e^{- \frac{\J \pi \left(\Re\{[\widehat{\ddot{\RV}}_{\XV}]_p\}+\Im\{[\widehat{\ddot{\RV}}_{\XV}]_q\}\right)}{2} } \bigg\}.
\end{align}\normalsize
Considering that $\widehat{\ddot{\RV}}_{\XV} \stackrel{D}{\rightarrow} {\cal CN}(\ddot{\RV}_{\XV}, \frac{4}{\pi^2 N}\SIGMAM)$, the expectations in \eqref{Eq-app-G-12} can be computed using the characteristic function of the Gaussian distribution as follows:
\small
\begin{align}
\label{Eq-app-G-13}
&\EX\{[\widetilde{\ddot{\RV}}]_p[\widetilde{\ddot{\RV}}]_q^*\} =  \frac{e^{\frac{-[\SIGMAM]_{p,p}-[\SIGMAM]_{q,q}}{4N}}}{2} \nonumber\\
&\times \bigg[ \cos\left(\frac{\pi}{2} \left[\Re\{[\ddot{\RV}_{\XV}]_p\}-\Re\{[\ddot{\RV}_{\XV}]_q\}\right] \right)  e^{\frac{\Re\{[\SIGMAM]_{p,q}\}}{2 N}} \nonumber\\
&- \cos\left(\frac{\pi}{2} \left[\Re\{[\ddot{\RV}_{\XV}]_p\}+\Re\{[\ddot{\RV}_{\XV}]_q\}\right] \right)  e^{-\frac{\Re\{[\SIGMAM]_{p,q}\}}{2 N}} \nonumber\\
& + \cos\left(\frac{\pi}{2} \left[\Im\{[\ddot{\RV}_{\XV}]_p\}-\Im\{[\ddot{\RV}_{\XV}]_q\}\right] \right)  e^{\frac{\Re\{[\SIGMAM]_{p,q}\}}{2 N}} \nonumber\\
&- \cos\left(\frac{\pi}{2} \left[\Im\{[\ddot{\RV}_{\XV}]_p\}+\Im\{[\ddot{\RV}_{\XV}]_q\}\right] \right)  e^{-\frac{\Re\{[\SIGMAM]_{p,q}\}}{2 N}}\nonumber\\
&+ \J \cos\left(\frac{\pi}{2} \left[\Im\{[\ddot{\RV}_{\XV}]_p\}-\Re\{[\ddot{\RV}_{\XV}]_q\}\right] \right)  e^{\frac{\Im\{[\SIGMAM]_{p,q}\}}{2 N}} \nonumber\\
&-\J \cos\left(\frac{\pi}{2} \left[\Im\{[\ddot{\RV}_{\XV}]_p\}+\Re\{[\ddot{\RV}_{\XV}]_q\}\right] \right)  e^{-\frac{\Im\{[\SIGMAM]_{p,q}\}}{2 N}} \nonumber\\
& - \J \cos\left(\frac{\pi}{2} \left[\Re\{[\ddot{\RV}_{\XV}]_p\}-\Im\{[\ddot{\RV}_{\XV}]_q\}\right] \right)  e^{-\frac{\Im\{[\SIGMAM]_{p,q}\}}{2 N}} \nonumber\\
&+ \J \cos\left(\frac{\pi}{2} \left[\Re\{[\ddot{\RV}_{\XV}]_p\}+\Im\{[\ddot{\RV}_{\XV}]_q\}\right] \right)  e^{\frac{\Im\{[\SIGMAM]_{p,q}\}}{2 N}} \bigg].
\end{align}\normalsize
Exploiting the Taylor expansion of the exponential function, \eqref{Eq-app-G-13} can be approximated for sufficiently large $N$ as
\small
\begin{align}
\label{Eq-app-G-14}
&\EX\{[\widetilde{\ddot{\RV}}]_p[\widetilde{\ddot{\RV}}]_q^*\} \simeq  \frac{1}{2} \\
\times&\bigg[ \cos\left(\frac{\pi}{2} \left[\Re\{[\ddot{\RV}_{\XV}]_p\}-\Re\{[\ddot{\RV}_{\XV}]_q\}\right] \right) \left(1+\frac{\Re\{[\SIGMAM]_{p,q}\}}{2 N}\right) \nonumber\\
&- \cos\left(\frac{\pi}{2} \left[\Re\{[\ddot{\RV}_{\XV}]_p\}+\Re\{[\ddot{\RV}_{\XV}]_q\}\right] \right)  \left(1-\frac{\Re\{[\SIGMAM]_{p,q}\}}{2 N}\right) \nonumber\\
& + \cos\left(\frac{\pi}{2} \left[\Im\{[\ddot{\RV}_{\XV}]_p\}-\Im\{[\ddot{\RV}_{\XV}]_q\}\right] \right)  \left( 1+ \frac{\Re\{[\SIGMAM]_{p,q}\}}{2 N}\right) \nonumber\\
&- \cos\left(\frac{\pi}{2} \left[\Im\{[\ddot{\RV}_{\XV}]_p\}+\Im\{[\ddot{\RV}_{\XV}]_q\}\right] \right)  \left(1-\frac{\Re\{[\SIGMAM]_{p,q}\}}{2 N}\right)\nonumber\\
&+ \J \cos\left(\frac{\pi}{2} \left[\Im\{[\ddot{\RV}_{\XV}]_p\}-\Re\{[\ddot{\RV}_{\XV}]_q\}\right] \right)  \left(1+\frac{\Im\{[\SIGMAM]_{p,q}\}}{2 N}\right) \nonumber
\end{align}
\begin{align}
&-\J \cos\left(\frac{\pi}{2} \left[\Im\{[\ddot{\RV}_{\XV}]_p\}+\Re\{[\ddot{\RV}_{\XV}]_q\}\right] \right)  \left(1-\frac{\Im\{[\SIGMAM]_{p,q}\}}{2 N}\right) \nonumber\\
& - \J \cos\left(\frac{\pi}{2} \left[\Re\{[\ddot{\RV}_{\XV}]_p\}-\Im\{[\ddot{\RV}_{\XV}]_q\}\right] \right)  \left(1-\frac{\Im\{[\SIGMAM]_{p,q}\}}{2 N}\right) \nonumber\\
&+ \J \cos\left(\frac{\pi}{2} \left[\Re\{[\ddot{\RV}_{\XV}]_p\}+\Im\{[\ddot{\RV}_{\XV}]_q\}\right] \right)  \left(1+\frac{\Im\{[\SIGMAM]_{p,q}\}}{2 N}\right) \bigg] \nonumber\\
&= \sin(\frac{\pi}{2} \Re\{[\ddot{\RV}_{\XV}]_p\}) \sin(\frac{\pi}{2} \Re\{[\ddot{\RV}_{\XV}]_p\}) \nonumber\\
&+ \sin(\frac{\pi}{2} \Im\{[\ddot{\RV}_{\XV}]_p\}) \sin(\frac{\pi}{2} \Im\{[\ddot{\RV}_{\XV}]_p\})
\nonumber\\
&+\J \sin(\frac{\pi}{2} \Im\{[\ddot{\RV}_{\XV}]_p\}) \sin(\frac{\pi}{2} \Re\{[\ddot{\RV}_{\XV}]_p\}) \nonumber\\
&-\J \sin(\frac{\pi}{2} \Re\{[\ddot{\RV}_{\XV}]_p\}) \sin(\frac{\pi}{2} \Im\{[\ddot{\RV}_{\XV}]_p\}) \nonumber\\
&+\frac{\Re\{[\SIGMAM]_{p,q}\}}{2N}\bigg[ \cos(\frac{\pi}{2} \Re\{[\ddot{\RV}_{\XV}]_p\}) \cos(\frac{\pi}{2} \Re\{[\ddot{\RV}_{\XV}]_p\}) \nonumber\\
&+ \cos(\frac{\pi}{2} \Im\{[\ddot{\RV}_{\XV}]_p\}) \cos(\frac{\pi}{2} \Im\{[\ddot{\RV}_{\XV}]_p\}) \bigg] \nonumber\\
&+\J +\frac{\Im\{[\SIGMAM]_{p,q}\}}{2N}\bigg[ \cos(\frac{\pi}{2} \Im\{[\ddot{\RV}_{\XV}]_p\}) \cos(\frac{\pi}{2} \Re\{[\ddot{\RV}_{\XV}]_p\}) \nonumber\\
&+ \cos(\frac{\pi}{2} \Re\{[\ddot{\RV}_{\XV}]_p\}) \cos(\frac{\pi}{2} \Im\{[\ddot{\RV}_{\XV}]_p\}) \bigg] .
\end{align}\normalsize
Consequently, it follows from \eqref{Eq-Est-2} that
\small
\begin{align}
\label{Eq-app-G-15}
[\GAMMAM]_{p,q}=&\EX\{[\widetilde{\ddot{\RV}}]_p[\widetilde{\ddot{\RV}}]_q^*\} \simeq [\ddot{\RV}]_p [\ddot{\RV}]_q^* \\
&+ \frac{1}{2N}  \bigg(\sqrt{1-[\Re\{[\ddot{\RV}]_p\}]^2} \times \sqrt{1-[\Re\{[\ddot{\RV}]_q\}]^2}\nonumber\\
&+ \sqrt{1-[\Im\{[\ddot{\RV}]_p\}]^2} \times \sqrt{1-[\Im\{[\ddot{\RV}]_q\}]^2}\bigg) \Re\{[\SIGMAM]_{p,q}\} \nonumber\\
&+\frac{\J}{2 N} \bigg(\sqrt{1-[\Im\{[\ddot{\RV}]_p\}]^2} \times \sqrt{1-[\Re\{[\ddot{\RV}]_q\}]^2} \nonumber\\
&+ \sqrt{1-[\Re\{[\ddot{\RV}]_p\}]^2} \times \sqrt{1-[\Im\{[\ddot{\RV}]_q\}]^2}\bigg) \Im\{[\SIGMAM]_{p,q}\} . \nonumber
\end{align}\normalsize
Inserting \eqref{Eq-app-G-15} into \eqref{Eq-app-G-11} and making use of \eqref{Eq-Est-5} yields
\small
\begin{align}
\label{Eq-app-G-16}
 &\EX\{\PSIM(\widehat{\PHI} \!-\! \PHI) (\widehat{\PHI} \!-\! \PHI)^H\PSIM^H\} \!\simeq\! \\
 &\left( \overline{\JM}^H \FM^H \DIAG(\BV) \FM^{-H} \SIGMAM^{-1} \FM^{-1} \DIAG(\BV) \FM \overline{\JM} \right)^{-1}\times \nonumber\\
& \overline{\JM}^H \FM^H \DIAG(\BV) \FM^{-H} \SIGMAM^{-1} \FM^{-1} \DIAG(\BV) \FM \GAMMAM \FM^H \DIAG(\BV) \FM^{-H} \SIGMAM^{-1}\FM^{-1}  \nonumber\\
&\times \DIAG(\BV) \FM \overline{\JM}\left( \overline{\JM}^H \FM^H \DIAG(\BV) \FM^{-H} \SIGMAM^{-1} \FM^{-1} \DIAG(\BV) \FM \overline{\JM} \right)^{-1}, \nonumber
\end{align}\normalsize
Finally, substituting \eqref{Eq-app-G-16} into \eqref{Eq-app-G-10} and considering that $\overline{p}_k = \frac{p_k}{\sigma^2 + \sum_{k=1}^K p_k}$ concludes the proof of Theorem \ref{Theo-6}.
\vspace{-2mm}
\section{Proof of Corollary \ref{Col-2}}
\label{app-I}
OCAB-MUSIC employs $\widetilde{\overline{\RV}} = \VE(\widetilde{\overline{\RM}})$ instead of $\widehat{\overline{\RV}}$. Hence, its asymptotic estimation error is obtained by replacing $\Delta \overline{\RV}$ with $\widetilde{\overline{\RV}} - \overline{\RV}$ in \eqref{Eq-app-G-1}. Following the same steps from \eqref{Eq-app-G-2} to \eqref{Eq-app-G-6}, the covariance of the asymptotic distribution (as $N \to \infty$) of the DoA estimation errors for OCAB-MUSIC is obtained as
\small
\begin{align}
\label{Eq-app-I-1}
{\cal E}_{\theta_{k_1}, \theta_{k_2}} =  \Re\{ \ZV^T_{k_1}  \TM \JM^{\dagger} \EX\{(\widetilde{\overline{\RV}} - \overline{\RV}) (\widetilde{\overline{\RV}} - \overline{\RV})^H \} \JM^{\dagger H} \TM^H \ZV_{k_2}^* \}.
\end{align}\normalsize
Considering the fact that the diagonal elements of $\widetilde{\overline{\RM}}$ and $\overline{\RM}$ are equal to one, \eqref{Eq-app-I-1} is simplified as
\small
\begin{align}
\label{Eq-app-I-2}
{\cal E}_{\theta_{k_1}, \theta_{k_2}} =  \Re\{ \ZV^T_{k_1}  \overline{\TM} \overline{\JM}^{\dagger} \left[\EX\{\widetilde{\ddot{\RV}}\widetilde{\ddot{\RV}}^H\} - \ddot{\RV} \ddot{\RV}^H \right] \overline{\JM}^{\dagger H} \overline{\TM}^H \ZV_{k_2}^* \}.
\end{align}\normalsize
Substituting $\EX\{\widetilde{\ddot{\RV}}\widetilde{\ddot{\RV}}^H\}$ from \eqref{Eq-app-G-15} completes the proof.
%
\section{Proof of Theorem \ref{Theo-7}}
\label{app-J}
To derive $\lim_{SNR \to \infty} {\cal E}_{\theta_k}$, we need to calculate $\lim_{SNR \to \infty}  (\sigma^2+\sum_{k'=1}^K p_{k'})^2/p_k^2$, $\WM_{\infty} = \lim_{SNR \to \infty} \WM$ and $\GAMMAM_{\infty} = \lim_{SNR \to \infty} \GAMMAM$. It is obtained from \eqref{Eq-app-E-1} that
\small
\begin{align}
\lim_{SNR \to \infty}  \frac{(\sigma^2+\sum_{k'=1}^K p_{k'})^2}{p_k^2} =  \lim_{SNR \to \infty} \frac{1}{\overline{p}_k^2} = K^2.
\end{align}\normalsize
In addition, it follows from \eqref{Eq-Est-22} and \eqref{Eq-Est-23} that $\WM$ and $\GAMMAM$ depend on SNR through $\overline{\RM}$, $\GAMMA$ and $\ddot{\RV}$. Hence, for calculating $\WM_{\infty}$ and $\GAMMAM_{\infty}$,  it is sufficient to first compute $\overline{\RM}_{\infty} = \lim_{SNR \to \infty} \overline{\RM}$, $\GAMMA_{\infty} = \lim_{SNR \to \infty} \GAMMA$ and $\ddot{\RV}_{\infty} = \lim_{SNR \to \infty} \ddot{\RV}$, and then insert them back into the expressions of $\WM$ and $\GAMMAM$ given in \eqref{Eq-Est-22} and \eqref{Eq-Est-23}. $\overline{\RM}_{\infty}$ is obtained in \eqref{Eq-app-E-2}. Given $\overline{\RM}_{\infty}$, $\GAMMA_{\infty} = \lim_{SNR \to \infty} \GAMMA$ is equal to the $(M^2-M) \times 1$ vector containing  the real and imaginary parts of the elements of $\overline{\RM}_{\infty}$ above its main diagonal elements. Then, exploiting \eqref{Eq-Est-10}, we have $\ddot{\RV}_{\infty} = \lim_{SNR \to \infty} \PSIM^{-1}\overline{\JM}^{\dagger}\FM^{-1} \GAMMA = \PSIM^{-1}\overline{\JM}^{\dagger}\FM^{-1} \GAMMA_{\infty}$. This completes the proof.
\bibliographystyle{IEEEtran}
\bibliography{Main}

\end{document}